\documentclass[twoside,leqno,twocolumn]{article}
\usepackage{ltexpprt}

\usepackage{bm}
\usepackage{algorithm}
\usepackage{multirow,caption}
\usepackage{verbatim}
\usepackage{algpseudocode}
\usepackage[english]{babel}
\usepackage[utf8]{inputenc}
\usepackage{cite}

\usepackage[resetlabels,labeled]{multibib}

\usepackage{booktabs}
\usepackage{etoolbox}
\makeatletter
\patchcmd{\@makecaption}
  {\scshape}
  {}
  {}
  {}
\makeatletter
\patchcmd{\@makecaption}
  {\\}
  {.\ }
  {}
  {}
\makeatother
\def\tablename{Table}

\newcounter{packednmbr}

\newenvironment{packeditemize}{\begin{list}{$\bullet$}{\setlength{\itemsep}{0pt}\addtolength{\labelwidth}{-5pt}\setlength{\leftmargin}{\labelwidth}\setlength{\listparindent}{\parindent}\setlength{\parsep}{0pt}\setlength{\topsep}{3pt}}}{\end{list}}

\newcommand{\normal}{\ensuremath{\mathtt{N}}}

\usepackage{cite}
\usepackage{amsmath,amssymb,amsfonts}
\usepackage{graphicx}
\usepackage{textcomp}
\usepackage{xcolor}
\def\BibTeX{{\rm B\kern-.05em{\sc i\kern-.025em b}\kern-.08em
    T\kern-.1667em\lower.7ex\hbox{E}\kern-.125emX}}

\algrenewcommand\alglinenumber[1]{%
  \scriptsize#1:}

\usepackage{hyperref}       
 
\usepackage{url}

\newcommand{\techrep}[2]{#2}

\makeatletter
\def\thanks#1{\protected@xdef\@thanks{\@thanks
        \protect\footnotetext{#1}}}
\makeatother

\begin{document}

\title{\Large Accelerated Experimental Design for Pairwise Comparisons}
\author{Yuan Guo$^{1}$\and Jennifer Dy$^{1}$ \and Deniz Erdogmus$^{1}$\and
Jayashree Kalpathy-Cramer$^{2}$ \and Susan Ostmo$^{3}$\and J. Peter Campbell$^{3}$\and Michael F. Chiang$^{3}$ \and Stratis Ioannidis$^{1}$   
\thanks{\scriptsize{$^1$\{yuanee20, erdogmus, jdy, ioannidis\}@ece.neu.edu. ECE Department, Northeastern University, Boston, MA, USA.
$^2$kalpathy@nmr.mgh.harvard.edu. Department of Radiology, Massachusetts General Hospital, Charlestown, MA, USA. $^3$\{ostmo, campbelp, chiangm\}@ohsu.edu Dept of Ophthalmology, Casey Eye Institute, Oregon Health $\&$ Science University, Portland, OR, USA.}}}

\date{}

\maketitle


\techrep{\fancyfoot[R]{\scriptsize{Copyright \textcopyright\ 2019 by SIAM\\Unauthorized reproduction of this article is prohibited}}}{}





\begin{abstract} 
Pairwise comparison labels are more informative and less variable than class labels, but generating them poses a challenge: their number grows quadratically in the dataset size. 
We study a natural experimental design objective, namely, D-optimality, that can be used to identify which $K$ pairwise comparisons to generate. This objective is known to perform well in practice, and is submodular, making the selection approximable via the greedy algorithm. A na\"ive greedy implementation has $O(N^2d^2K)$ complexity, where $N$ is the dataset size,  $d$ is the feature space dimension, and $K$ is the number of  generated comparisons. We show that, by exploiting the inherent geometry of the dataset--namely, that it consists of pairwise comparisons--the greedy algorithm's complexity can be reduced to $O(N^2(K+d)+N(dK+d^2) +d^2K).$ We apply the same acceleration also to the so-called lazy greedy algorithm. When combined, the above improvements lead to an execution time of less than 1 hour for a dataset with $10^8$ comparisons; the na\"ive greedy algorithm on the same dataset would require more than 10 days to terminate.
\end{abstract}

\section{Introduction}
 In many supervised learning applications, including medicine and recommender systems, class labels are solicited from (and generated by) human labelers. Datasets constructed thusly are often noisy, 
 to counter this, several recent works \cite{kalpathy2016plus,sculley2010combined,desarkar2010aggregating, desarkar2012preference,guo} propose augmenting datasets via \emph{comparisons}. For example, a medical expert can classify patients as, e.g. diseased or normal, but can also order pairs of patients w.r.t. disease severity.  Similarly, beyond generating class labels in recommender systems  (e.g., stars), labelers can also declare their relative preference between any two items.

Incorporating comparison labels to the training process has two advantages. First,  comparisons indeed reveal additional information compared to traditional class labels: this is because they capture  both inter \emph{and} intra-class relationships; the latter are not revealed via class labels alone.
Second, comparisons are often less noisy than (absolute) class labels. Indeed, human labelers disagreeing when generating class judgments often exhibit reduced variability when asked to compare pairs of items instead. This has been extensively documented in a broad array of domains, including medicine \cite{stewart2005absolute, kalpathy2016plus}, movie recommendations \cite{brun2010towards, desarkar2010aggregating}, travel recommendations \cite{zheng2009mining}, music recommendations \cite{koren2011ordrec}, and web page recommendations \cite{schultz2004learning}, to name a few.


Nevertheless, soliciting comparison labels poses a significant challenge, as the number of potential comparisons is quadratic in the dataset size. It therefore makes sense to solve the following experimental design (i.e., batch active learning) problem:\emph{ given a budget  $K$, and a set of existing class labels, identify the $K$ comparison labels the expert should generate that will better augment the existing dataset}. There are several natural ways through which this experimental design problem can be formalized. In this paper, we focus on an objective motivated by D-Optimal design \cite{boyd2004convex,pukelsheim1993optimal}. This objective leads to selections that perform very well in practice against competing methods \cite{guo}. Most importantly, it is also \emph{submodular}; as such, a set of comparisons attaining a constant approximation guarantee can be constructed in polynomial time via the so-called \emph{greedy algorithm} \cite{nemhauser1978}.

Applying the greedy algorithm na\"ively in this  experimental design setting leaves a lot to be desired. Given that the set of comparisons is quadratic, a na\"ive implementation of the algorithm leads to a complexity of $O(N^2d^2K)$, where $N$ is the size of the dataset, $d$ is the dimension of the feature space, and $K$ is the size of the selected set of comparisons. The quadratic nature of this computation makes the use of the algorithm prohibitive for all but the smallest datasets, especially when the samples are high-dimensional. On the other hand, the fact that the same $N$ objects  participate in these $O(N^2)$ pairs suggests an underlying structure that can potentially be exploited to improve time performance.   

To that end, we make the following contributions:
\begin{packeditemize}
\item We formally study the problem of accelerating the greedy algorithm for learning pairwise comparisons. To the best of our knowledge, we are the first to study methods of reducing the complexity of greedy by exploiting the inherent geometry of the dataset-namely, that it consists of pairwise comparisons.
 
\item We show that, by exploiting this underlying structure, the greedy algorithm can indeed be accelerated. Using Cholesky factorization \cite{van1996matrix}, the Sherman Morisson formula\cite{sherman1950adjustment}, and the pairwise comparison structure, we reduce the   greedy algorithm's complexity from $O(N^2d^2K)$ to $O(N^{2}(K+d)+N(dK+d^{2})+d^{2}K)$. The  $O(N^2(K+d))$ term, which dominates when $N\gg K+d$, consists of an $O(N^2d)$ pre-processing step and an $O(N^2)$ computation per iteration involving only scalar operations. 
 
 \item We further apply  our acceleration techniques to the so-called \emph{lazy-greedy} algorithm \cite{minoux1978accelerated,mirzasoleiman2015lazier,lin2011class,mirzasoleiman2013distributed}, which is  known to perform well experimentally.
   \item We  evaluate  the execution time performance of our accelerated algorithms over both  synthetic  and  real-life  datasets, demonstrating that they significantly outperform na\"ive implementations. Our experiments show that we can select comparisons from a dataset involving  more than $10^{8}$ comparison pairs, each comprising $400$-dimensional features, in less than an hour; a na\"ive implementation  takes more than $10$ days. 
\end{packeditemize}
The remainder of this paper is organized as follows. We discuss related work in Section \ref{sec:related}. Our problem formulation and our accelerated greedy  algorithm can be found in Sections \ref{sec:Problem} and \ref{sec:aca}, respectively. We discuss our accelerated lazy greedy algorithms in Sec.~\ref{sec:aclazy}, and present our numerical evaluations  in Section~\ref{sec:evaluation}. Finally, we conclude in Section~\ref{sec:conclusion}.
\section{Related Work}\label{sec:related}




Integrating classification and pairwise comparison labels has received considerable attention recently \cite{sculley2010combined, chen2015fusing, takamura2015estimating,wang2016ppp}. 
 Integrating regression labels with ranking information was proposed in \cite{sculley2010combined} as a means to improve regression outcomes in label-imbalanced datasets, and similar approaches have been used to incorporate both ``pointwise'' and ``pairwise'' labels in image classification tasks \cite{chen2015fusing,wang2016ppp}. Penalties used in this literature are variants of the MAP estimation we describe in Sec.~\ref{sec:Problem}, and are directly related to our Bradley-Terry generative model. None of these works however deal with the  problem of  how to collect pairwise comparison  labels. 

Experimental design (a.k.a.~batch active learning) is classic~\cite{pukelsheim1993optimal}. Mutual information is a a commonly used objective \cite{liepe2013maximizing,cavagnaro2010adaptive}, which is monotone submodular under certain conditions \cite{krause2012near}. Applying this objective to our generative model retains submodularity but, as in other settings \cite{busetto2013near}, both (a) computing the posterior of the model, as well as (b) evaluating the function when having access to this posterior, are intractable. Many natural objectives 
are submodular, and are thus amenable to approximation via the greedy algorithm by Nemhauser et al.~\cite{nemhauser1978}; indeed, submodularity  arises in a broad array of active learning problems \cite{krause2014submodular,golovin2011adaptive}.  

Our setting is closest to--and motivated by--work by a series of papers that study experimental design in the context of comparisons. Jamieson and Nowak \cite{active} assume the existence of a total ordering, and which is learned in the absence of features. 
Grasshof et al.~\cite{grasshoff2008optimal} and  Glickman et al. \cite{glickman2005adaptive}  study experimental design on the Bradley-Terry model, again without features. They use D-Optimal design  and KL-divergence  as optimization objectives, respectively. Closer to our setting, Guo et al.~\cite{guo} study four different submodular experimental design objectives, including D-optimality, Mutual Information, Information Entropy, and Fisher Information, in the high-dimensional setting. The authors establish experimentally that Mutual Information performs best, but is intractable, while D-optimality is a close second. Guo et al.~implement only the na\"ive greedy algorithm, whose complexity is  $O(N^2d^2K)$, and do not exploit the underlying structure of the problem to accelerate the algorithm; this limits their experiments to datasets with no more than $10^4$ comparisons. We depart in identifying ways to exploit this structure to drastically accelerate the greedy algorithm, enabling us to solve problems with $10^8$ comparisons in less than an hour.

\section{Problem Formulation}\label{sec:Problem}
\begin{table}[!t]

\begin{center}
{\scriptsize
\caption{Summary of Notation}\label{tab:notation}
\vspace*{-1em}
\begin{tabular}{c|l}
\hline
\!\!$\mathbb{N}$, $\mathbb{R}$, $\mathbb{S}_{+}$\!\!&  sets of naturals, reals, and positive definite matrices\\
$N$ & number of samples in dataset\\
    $d$& sample dimension (i.e., number of features)\\
$\mathcal{N}$ &  dataset of samples \\
$i,j$   &  sample indices in $\mathcal{N}$ \\
$\mathcal{C}$&  set of  pairwise comparisons \\
$\mathcal{A}$& the initial set with absolute labels\\
    $K$& number of comparisons to be collected \\
   $\mathcal{S}$ & subset  of $\mathcal{C}$ to be collected\\
       $\lambda$ & regularization parameter in $\mathbb{R}_+$\\
    $\bm{x}_i$ &  feature vector of sample $i$ in $\mathbb{R}^d$\\
    $\bm{x}_{i,j}$ & $\bm{x}_{i}-\bm{x}_{j}$\\
        $\mathbf{X}$&   matrix of feature vectors $\bm{x}_i$, $i\in \mathcal{N}$\\
        $\mathbf{A}$ & matrix used in D-optimality criterion, given by \eqref{eq:design} \\ 
        $f$ & submodular objective\\
        $y_i$ & absolute label of sample $i$\\
    $y_{i,j}$ & comparison outcome between $i$ and $j$ \\
        $s_i$ & Bradley-Terry score for sample $i$\\
            $\bm{\beta}$ & parameter vector/model in $\mathbb{R}^d$\\
$\Omega$ & abstract set in submodular maximization (for us, $\Omega=\mathcal{C}$) \\
    $e$ & abstract element in $\Omega$ (for us, $e=(i,j)\in\mathcal{C}$)\\
    $\bm{x}_{e}$ &$\bm{x}_{i}-\bm{x}_{j}$, where $e=(i,j)$\\
    $\Delta(e|\mathcal{S})$& marginal gain $f(\mathcal{S}\cup e)-f(\mathcal{S})$\\
    $d_{e}$& proxy for marginal gain of element $e$\\
    $\mathbf{U}$& Cholesky factor of matrix $\mathbf{A}^{-1}=\mathbf{U}^{T}\mathbf{U}$\\
    $\bm{z}_{i}$& vectors used in Factorization Greedy, equal to $\mathbf{U}\bm{x}_{i}$\\
    $\bm{v}$& auxiliary vector used in Scalar Greedy\\
    \hline
    \end{tabular}
    }
    \end{center}
    \vskip -0.12in
\end{table}

Consider a setting in which data samples are labeled by an expert. Given a sample to label, the expert produces a binary \emph{absolute} label, indicating the sample's class. Given two different samples, the expert produces a \emph{comparison} label. Comparison labels are also binary  and indicate  precedence with respect to the classification outcome. For example, for a medical diagnosis problem, absolute labels indicate the existence of disease, while  comparison labels indicate the relative severity between two samples. 
An experimenter has access to noisy absolute labels generated by this expert. At the same time, the experimenter wishes to augment the dataset by adding comparison labels. As comparison labels are numerous (quadratic in the dataset size) and their acquisition is time-consuming,  the experimenter collects only a subset of all possible comparison labels.

Formally, the experimenter has access to $N$ samples, indexed by $i\in\mathcal{N}\equiv \{1,\ldots,N\}$.  Every sample has a feature vector $\bm{x}_i\in\mathbb{R}^d$, known to the experimenter; we denote by $\mathbf{X}=[\mathbf{x}_i]_{i\in \mathcal{N}}\in \mathbb{R}^{n\times d}$ the matrix of feature vectors. For some set $\mathcal{A}\subseteq \mathcal{N}$, the experimenter has access to binary absolute labels $y_{i}\in\{+1,-1\}$, $i\in \mathcal{A}$,  generated by the expert. We define $\mathcal{C}\equiv\{(i,j):i,j\in \mathcal{N},i<j\}$ to be the set of possible pairwise comparisons. 

\subsection{Experimental Design.}
The experimenter wishes to augment the existing dataset of absolute labels by adding comparison labels $y_{i,j}\in\{+1,-1\}$, where $(i,j)\in\mathcal{C}$. It is expensive and time consuming to collect all
$|\mathcal{C}|=\frac{N(N-1)}{2}$ comparison labels. The experimenter thus collects $K$ labels from a subset $\mathcal{S}\subseteq\mathcal{C}$,  where $|\mathcal{S}|=K$. 
To determine the optimal such set $\mathcal{S}^*$, the experimenter solves: 
\begin{align}\label{eq:obj}
\begin{split}
\text{Maximize}~~& f(\mathcal{S})-f(\emptyset),\\
\text{subj.~to}~~&\mathcal{S}\subseteq \mathcal{C}, |\mathcal{S}|=K.
\end{split}
\end{align}
where  objective  $f:2^{|\mathcal{C}|}\rightarrow \mathbb{R}$ captures how informative samples in $\mathcal{S}$ are.
We use the objective: 
\begin{align}\label{eq:dobj}
 f(\mathcal{S})&=\mathop{\log\det}(\lambda \mathbf{I}_{d}\!+\!\sum_{i\in \mathcal{A}}\!\bm{x}_{i}\bm{x}_{i}^{T}\!+\!\sum_{(i,j)\in \mathcal{S}}\!\!\bm{x}_{i,j}\bm{x}_{i,j}^{T})
\end{align}
where $\bm{x}_{i,j}=\bm{x}_{i}-\bm{x}_{j}$, $\lambda>0$ is a positive value, and $I_{d}\in \mathbb{R}^{d\times d}$ is the $d$-dimensional identity matrix. As above, sets $\mathcal{A}$ and $\mathcal{S}$ represent the set of absolute labels observed already and the set of comparisons to be collected, respectively.

Objective \eqref{eq:dobj} is motivated by D-optimal design~\cite{boyd2004convex}, assuming a Bradley-Terry generative model for comparison labels \cite{bradley1952rank}. In particular,
\eqref{eq:dobj} is the negative log entropy of a linear model learned under Gaussian noise~\cite{boyd2004convex}, and has been observed to have excellent performance as an experimental design objective compared to a broad array of competitors, including Mutual Information and Fisher Information \cite{guo,he2010laplacian}. 
Before elaborating on how to solve \eqref{eq:obj}, we  briefly discuss  how \eqref{eq:dobj} arises under the Bradley-Terry model below. 

\subsection{ D-Optimal Design Under the Bradley-Terry Model.}\label{sec:gm}
Assume that absolute and comparison labels are generated according to the following  probabilistic  model. First, there exists a parameter vector $\bm{\beta}\in\mathbb{R}^{d}$, sampled from a Gaussian prior $\normal(0,\sigma^2\mathbf{I})$, such that for all $i\in\mathcal{N}$ and all $(i,j)\in\mathcal{C}$ the absolute labels $y_{i}$ and comparison labels $y_{i,j}$ are independent conditioned on $\bm{\beta}$. Second, the conditional distribution of $y_{i}$ given $\bm{x}_{i}$ and $\bm{\beta}$ is given by a logistic model, i.e.,
\begin{align}\label{eq1}
    \textstyle\mathbf{P}(y_{i}=+1|\bm{x}_{i},\bm{\beta})=\frac{1}{1+\exp(-\bm{\beta}^{T}\bm{x}_{i})},   \quad i\in \mathcal{N}.
\end{align}
Finally, the conditional distribution of $y_{i,j}$ given $\bm{x}_{i},\bm{x}_{j}$ and $\bm{\beta}$ is given by the following Bradley-Terry model \cite{bradley1952rank}: every sample $i\in \mathcal{N}$ is associated with a parameter $s(\bm{x}_{i},\bm{\beta})=\exp(\bm{\beta}^{T}\bm{x}_{i})\in \mathbb{R}_{+}$ such that, for all $(i,j)\in\mathcal{C}$,
\begin{align}\label{eq2}
\begin{split}
    \textstyle\mathbf{P}(y_{i,j}\!=\!+1|\bm{x}_{i},\bm{x}_{j},\bm{\beta})&=\textstyle\frac{s(\bm{x}_{i},\bm{\beta})}{s(\bm{x}_{i},\bm{\beta})+s(\bm{x}_{j},\bm{\beta})}
    \end{split}
\end{align}
%
%
Intuitively, score $s(\bm{x},\bm{\beta})$ captures the propensity of input $\bm{x}$ to receive a positive absolute label, \emph{as well as}  to be selected when compared to other objects.

The advantage of the generative model  \eqref{eq1}-\eqref{eq2} is that it leads to a tractable Maximum A-Posteriori (MAP) estimation procedure for learning $\bm{\beta}$. Indeed, after collecting both  absolute \emph{and} comparison labels, 
the experimenter learns $\bm{\beta}$ 
by minimizing the following negative log-likelihood loss function:
\begin{equation}\label{eq4}
\begin{split}
\mathcal{L}(\bm{\beta};\mathcal{A},\mathcal{S}) =&\textstyle \lambda ||\bm{\beta}||_{2}^{2} +\sum_{i\in \mathcal{A}}\mathbf{log}(1+e^{-y_i\bm{\beta}^T\bm{x_i}})
\\
\textstyle+\sum_{(i,j)\in \mathcal{S}}&\mathbf{log}(1+e^{-y_{i,j}\bm{\beta}^T(\bm{x_i}-\bm{x_j})}),
\end{split}
\end{equation}
where the coefficient $\lambda$ equals $1/\sigma^{2}$. 
The loss $\mathcal{L}(\bm{\beta};\mathcal{A},\mathcal{S}) $ is convex in $\bm{\beta}$; in fact, it can be seen as a special form of logistic regression, in which the covariates of comparison labels $y_{i,j}$ are given precisely by $\bm{x}_{i,j}=\bm{x}_{i}-\bm{x}_{j}.$
Matrix 
\begin{align}\label{eq:design}\textstyle\mathbf{A}(\mathcal{S}) = \lambda \mathbf{I}_{d}+\sum_{i\in \mathcal{A}}\bm{x}_{i}\bm{x}_{i}^{T}+\sum_{(i,j)\in \mathcal{S}}\bm{x}_{i,j}\bm{x}_{i,j}^{T}\end{align}
used in our objective \eqref{eq:dobj} is the Fisher information matrix resulting from \eqref{eq4}, when the underlying logistic regression is approximated by a linear regression. 

\subsection{Greedy Optimization.}
\begin{algorithm}[!t]
\begin{scriptsize}
\caption{Greedy Algorithm} \label{alg:greedy}
\begin{algorithmic}[1]
\Procedure{Greedy}{$f,\Omega$}
\State \textsc{PreProcessing}(\,)
	\While{$|\mathcal{S}|<K$}
	\State
	$e^{*}=\textsc{FindMax}(\mathcal{S})$	
	\State 
	$\textsc{UpdateS}(\mathcal{S},e^{*})$
	\EndWhile
	\State \Return $\mathcal{S}$
\EndProcedure
\end{algorithmic}
\begin{algorithmic}[1]
\Procedure {PreProcessing}{\,}
\State Set $\mathcal{S}=\emptyset$  
\EndProcedure
\end{algorithmic}
\begin{algorithmic}[1]
\Procedure {FindMax}{$\mathcal{S}$}
\State \Return $e^{*}=\underset{e\in \Omega\setminus\mathcal{S}}{\mathbf{argmax}}~  \Delta(e|\mathcal{S})$ \Comment{$\Delta(e|\mathcal{S})$  given by
\eqref{eq:marginal}.} 
\EndProcedure
\end{algorithmic}
\begin{algorithmic}[1]
\Procedure{UpdateS}{$\mathcal{S},e^{*}$}
\State Set $\mathcal{S}=\mathcal{S}\cup e^{*}$
\EndProcedure
\end{algorithmic}
\end{scriptsize}
\end{algorithm}

Unfortunately, problem \eqref{eq:obj} is NP hard both for the D-optimality objective \eqref{eq:dobj} as well as for many other objective functions of interest \cite{guo,krause2012near,krause2014submodular}. However, we can produce an approximation algorithm using the theory of submodular functions. A set function  $f: 2^{\Omega}\rightarrow \mathbb{R}$ is  \emph{submodular} if
$f(\mathcal{T}\cup \{z\})-f(\mathcal{T})\geqslant f(\mathcal{D}\cup \{z\})-f(\mathcal{D})$ for all $ \mathcal{T} \subseteq \mathcal{D} \subseteq \Omega$ and   $z\in\Omega $. Function $f$ is called  \emph{monotone} if $f(\mathcal{D}\cup \{z\})-f(\mathcal{D})\geqslant0$ for all $\mathcal{D} \subseteq \Omega$ and $z\in \Omega$. The \emph{greedy} algorithm, summarized in Alg.~\ref{alg:greedy}, solves problem
\begin{small}
\begin{align}\label{eq:gobj}
\begin{split}
\text{Maximize}~~& f(\mathcal{S}),\\
   \text{s.t.}~~&|\mathcal{S}|\leqslant K,\mathcal{S}\subseteq \Omega,  
\end{split}
\end{align}
\end{small}
where $f$ is monotone submodular over set $\Omega$. 
Starting from $\mathcal{S}=\emptyset$, the algorithm iteratively adds the element $e$ to the present set $\mathcal{S}$ that maximizes the  \emph{marginal gain}:
\begin{small}
\begin{align}\label{eq:marginal}
\Delta(e|\mathcal{S})=f(\mathcal{S}\cup e)-f(\mathcal{S}),  
\end{align}
\end{small}
among all elements $e\in\Omega\setminus\mathcal{S}$; this is repeated
until $|\mathcal{S}|=K$.   The following guarantee holds:
\begin{theorem} Nemhauser et al.~\cite{nemhauser1978}
If $f$ is monotone submodular,  the set $\mathcal{S}$ returned by Alg.~\ref{alg:greedy} satisfies:  
 $   f(\mathcal{S})-f(\emptyset)\geqslant(1-1/e) (f(\mathcal{S}^{*})-f(\emptyset)),$
where $\mathcal{S}^{*}$ is the optimal solution to Eq. \eqref{eq:gobj}.
\end{theorem}

Objective   \eqref{eq:dobj} is indeed monotone submodular. However, set $\Omega=\mathcal{C}$ is \emph{quadratic} in the number of inputs $N$. This is prohibitive for large datasets, particularly when  marginal gains $\Delta(e|\mathcal{S})$ are themselves expensive to compute.  As described  in the next section, for $f$ given by \eqref{eq:dobj}, each marginal gain computation is $O(d^{2})$; this motivates us to accelerate Alg.~\ref{alg:greedy}.

\section{Accelerating The Greedy Algorithm}\label{sec:aca}

\begin{algorithm}[!t]
\begin{scriptsize}
\caption{Na\"ive Greedy}\label{alg:ng}
Na\"ive greedy algorithm, as described in Sec.~\ref{subsec:naive}. The main \textsc{Greedy} procedure is the same as in Alg.~\ref{alg:greedy}.
\begin{algorithmic}[1]
\Procedure{Preprocessing}{$\mathbf{X}$}
\State Compute
	$\mathbf{A}_{0}^{-1}=(\lambda I_{d}+\underset{i\in\mathcal{A}}{\sum}\bm{x}_{i}\bm{x}_{i}^{T})^{-1}$; Set $\mathbf{A}^{-1}=\mathbf{A}_0^{-1}$; Set $\mathcal{S}=\emptyset$	
\EndProcedure
\end{algorithmic}
\begin{algorithmic}[1]
\Procedure {FindMax}{$\mathcal{S}$}
\State Compute $d_{e}=\bm{x}_{e}\mathbf{A}^{-1}\bm{x}_{e}$ for $e \in \mathcal{C} \setminus \mathcal{S}$
\State \Return $e^{*}=\underset{e\in \mathcal{C}\setminus\mathcal{S}}{\mathbf{argmax}}~~  d_{e}$
\EndProcedure
\end{algorithmic}
\begin{algorithmic}[1]
\Procedure{UpdateS}{$\mathcal{S},e^{*}$}
\State Set $\mathcal{S}=\mathcal{S}\cup e^{*}$;  Set $\mathbf{A}^{-1}=\mathbf{A}^{-1}-\frac{\mathbf{A}^{-1}\bm{x}_{e^{*}}\bm{x}_{e^{*}}^{T}\mathbf{A}^{-1}}{1+\bm{x}_{e^{*}}^{T}\mathbf{A}^{-1}\bm{x}_{e^{*}}}$
\EndProcedure
\end{algorithmic}

\end{scriptsize}
\end{algorithm}

In this section, we describe how to accelerate the greedy algorithm, improving its complexity from $O(N^{2}d^{2}K)$ to $O(N^{2}(d+K)+N(dK+d^{2})+Kd^{2})$. In doing so, we exploit the inherent structure of set $\Omega=\mathcal{C}$, namely, that it comprises pairwise comparisons. When $N\gg d+K$, the dominant term is $O(N^2(d+K))$; constants in this term amount to the time to compute 1 scalar multiplication and 1 scalar addition; as such, the algorithm scales very well in practice (see Sec.~\ref{sec:evaluation}). 

Before presenting our accelerated method, we first review a na\"ive implementation (\emph{Na\"ive Greedy}) of Alg.~\ref{alg:greedy} applied to our problem \eqref{eq:obj}.  We also construct an intermediate algorithm (\emph{Factorization Greedy}), with slightly improved complexity ($O(Nd^{2}K+N^{2}dK)$) over Na\"ive Greedy. Finally, we present our fastest algorithm (\emph{Scalar Greedy}), that attains the aforementioned guarantee. We present Factorization Greedy both for the sake of clarity, but also because its lazy implementation, presented in Section~\ref{sec:aclazy}, has advantages over the corresponding Scalar Greedy algorithm. All algorithms receive the sample feature matrix $\mathbf{X}\in\mathbb{R}^{n\times d}$ as input.  

\subsection{Na\"ive Greedy.}\label{subsec:naive}
Our first ``na\"ive'' implementation slightly improves upon the abstract greedy algorithm (Alg.~\ref{alg:greedy}), which operates on the value oracle model, by (a) computing a simpler version of gains $\Delta(e|\mathcal{S})$, and (b) speeding-up matrix inversion via the Sherman-Morisson formula \cite{sherman1950adjustment}. For $f$ given by Eq.~\eqref{eq:dobj}, by the matrix determinant lemma\cite{harville1997matrix}:
\begin{small}
\begin{align}
\begin{split}
\Delta(e|\mathcal{S})&=
\mathop{\log}(1+\bm{x}_{e}^{T}\mathbf{A}^{-1}\bm{x}_{e}),
\end{split}
\end{align}
\end{small}
where $\mathbf{A}=\mathbf{A}(\mathcal{S})\equiv(\lambda \mathbf{I}_{d}+\sum_{i\in \mathcal{A}}\bm{x}_{i}\bm{x}_{i}^{T}+\sum_{e\in \mathcal{S}}\bm{x}_{e}\bm{x}_{e}^{T})\in \mathbb{S}_{+}^{d}$ and $\bm{x}_{e}= \bm{x}_{i,j}=\bm{x}_{i}-\bm{x}_{j}$ for all  $e\equiv (i,j)\in \mathcal{C}\setminus\mathcal{S}$.  As $\mathop{\log}(1+s)$ is monotone in $s$, to implement \textsc{FindMax} in Alg.~\ref{alg:greedy}, it suffices to compute the maximum among
\begin{small}
\begin{align}\label{eq:de}
    d_{e}=d_e(S)\equiv\bm{x}_{e}^{T}\mathbf{A}^{-1}\bm{x}_{e},~~~ e=(i,j)\in \mathcal{C}\setminus \mathcal{S}.
\end{align}
\end{small}
We call $d_e$, $e\in \mathcal{C}$, the  \emph{proxy} marginal gain. We further reduce computation costs using the fact that 
\begin{small}
\begin{align}\label{eq:sherman}
  \textstyle  \mathbf{A}^{-1}(\mathcal{S}\cup e^{*})=
    \mathbf{A}^{-1}(\mathcal{S})-\frac{\mathbf{A}^{-1}(\mathcal{S})\bm{x}_{e^{*}}\bm{x}_{e^{*}}^{T}\mathbf{A}^{-1}(\mathcal{S})}{1+\bm{x}_{e^{*}}^{T}\mathbf{A}^{-1}(\mathcal{S})\bm{x}_{e^{*}}},
\end{align}
\end{small}
by the Sherman Morrison formula \cite{sherman1950adjustment}. These two observations lead to the implementation of the na\"ive greedy algorithm presented in Alg.~\ref{alg:ng}. The algorithm uses the same main Greedy procedure as Alg.~\ref{alg:greedy}. In preprocessing, we initialize matrix $\mathbf{A}^{-1}$. At each iteration, we find element $e^{*}$ that maximizes $d_{e}$ rather than $\Delta(e|\mathcal{S})$, and subsequently update $\mathbf{A}^{-1}$ via \eqref{eq:sherman}.

Inverting matrix $\mathbf{A}_0$ has complexity\footnote{As matrix inversion has the same complexity as matrix multiplication.} $O(d^{2.37})$, though for small $\mathcal{A}$ the Sherman-Morisson formula can be used again to reduce this to $O(d^2|\mathcal{A}|)$.
Computing $d_{e}$ and updating $\mathbf{A}^{-1}$ via the Sherman Morrison formula have complexity $O(d^{2})$.   Hence,  Alg.~\ref{alg:ng} has a total complexity $O(N^{2}d^{2}K)$, which scales poorly for high $N$ and $d$. Note that the $O(d^{2.37})$ term in pre-processing is dominated by higher order terms and therefore ignored; this holds for all algorithms in this section.  




\subsection{Factorization Greedy.}\label{subsec:fga}
\begin{algorithm}[!t]
\begin{scriptsize}
\caption{Factorization Greedy}\label{alg:fga}
Factorization greedy algorithm, as described in Sec.~\ref{subsec:fga}. The main \textsc{Greedy} procedure is the same as in Alg.~\ref{alg:greedy}, while the  \textsc{PreProcessing} and \textsc{UpdateS} are the same as Alg.~\ref{alg:ng}.
\begin{algorithmic}[1]
\Procedure {FindMax}{$\mathcal{S}$}
\State Factorize the matrix $\mathbf{A}^{-1}$ into $\mathbf{A}^{-1}=\mathbf{U}^{T}\mathbf{U}$ by Cholesky factorization.
	
\State Compute and save  $\bm{z}_{i}=\mathbf{U}\bm{x}_{i}$ for all $i\in \mathcal{N}$. \label{line:zi1} 
\State Compute and save  $d_{e}=||\bm{z}_{i}-\bm{z}_{j}||_{2}^{2}$ for all $e\in \mathcal{C}\setminus\mathcal{S}$.
\State \Return $e^{*}=\underset{e\in \mathcal{C}\setminus\mathcal{S}}{\mathbf{argmax}}~~  d_{e}$
\EndProcedure
\end{algorithmic}
\end{scriptsize}
\end{algorithm}

Na\"ive Greedy requires $O(N^{2}d^{2})$ operations per iteration. To avoid this, we  exploit the pairwise comparison structure of $\bm{x}_{e}=\bm{x}_{i}-\bm{x}_{j}$, for $e=(i,j)\in \mathcal{C}$. Note that positive definite matrix $\mathbf{A}^{-1}$ can be factorized into  $\mathbf{A}^{-1}=\mathbf{U}^{T}\mathbf{U}$ by Cholesky factorization 
, where matrix $\mathbf{U}$ is an upper triangular matrix. Then, $d_{e}$ satisfies:
\begin{small}
\begin{align}\label{eq:factor}
d_{e}&=\bm{x}_{e}^{T}\mathbf{A}^{-1}\bm{x}_{e}
=||\mathbf{U}\bm{x}_{e}||_{2}^{2}=
||\mathbf{U}\bm{x}_{i}-\mathbf{U}\bm{x}_{j}||_{2}^{2}.
\end{align} 
\end{small}
This gives rise to the following algorithm, summarized in Alg.~\ref{alg:fga}. 
\textsc{PreProcessing} and \textsc{UpdateS} are as in the Na\"ive Greedy algorithm (Alg.~\ref{alg:ng}). For \textsc{FindMax}, in each iteration, we first factorize the matrix $\mathbf{A}^{-1}$ into $\mathbf{U}^{T}\mathbf{U}$ and calculate and save $\bm{z}_{i}=\mathbf{U}\bm{x}_{i}$ for all $i\in \mathcal{N}$. Then we calculate $d_{e}$ via Eq.~\eqref{eq:factor} for all $e\in \mathcal{C}\setminus\mathcal{S}$, and return the maximal element.
Cholesky factorization has $O(d^{2.37})$ complexity \cite{press2007numerical}. Computing $\mathbf{U}\bm{x}_{i}$ for all $i\in\mathcal{N}$ involves $O(Nd^{2})$ computations, while computing all $d_{e}$, $e\in \mathcal{C}\setminus\mathcal{S}$, via Eq.~\eqref{eq:factor} requires $O(N^{2}d)$ computations. 
Hence, the complexity of \textsc{FindMax} in Alg.~\ref{alg:fga} is  $O(Nd^{2}+N^{2}d)$, and the entire Factorization Greedy algorithm has  complexity $O(Nd^{2}K+N^{2}dK)$.

\subsection{Scalar Greedy.}\label{sec:Scalar}

\begin{algorithm}[!t]
\begin{scriptsize}
\caption{Scalar Greedy}\label{alg:sg} Scalar greedy algorithm, as described in Sec.~\ref{sec:Scalar}. The main \textsc{Greedy} procedure is the same as in Alg.~\ref{alg:greedy}.
\begin{algorithmic}[1]
\Procedure{Preprocessing}{$\mathbf{X}$}
\State Compute
	$\mathbf{A}_{0}^{-1}=(\lambda I_{d}+\underset{i\in\mathcal{A}}{\sum}\bm{x}_{i}\bm{x}_{i}^{T})^{-1}$; Set $\mathbf{A}^{-1}=\mathbf{A}_0^{-1}$; Set $\mathcal{S}=\emptyset$.
\State Compute $\mathbf{U}$: factorize the matrix $\mathbf{A}_{0}^{-1}$ into $\mathbf{U}^{T}\mathbf{U}$ by Cholesky factorization.
	
\State Compute and save  $\bm{z}_{i}=\mathbf{U}\bm{x}_{i}$ for all $i\in \mathcal{N}$. 
\State Compute and save  $d_{e}=||\bm{z}_{i}-\bm{z}_{j}||_{2}^{2}$ for all $e\in \mathcal{C}$.
\EndProcedure
\end{algorithmic}
\begin{algorithmic}[1]
\Procedure {FindMax}{$\mathcal{S}$}
\State \Return $e^{*}=\underset{e\in \mathcal{C}\setminus\mathcal{S}}{\mathbf{argmax}}~~  d_{e}$
\EndProcedure
\end{algorithmic}
\begin{algorithmic}[1]
\Procedure{UpdateS}{$\mathcal{S},e^{*}$}
\State $\mathcal{S}=\mathcal{S}\cup e^{*}$
\State Compute $\bm{v}=\frac{\mathbf{A}^{-1}\bm{x}_{e^{*}}}{\sqrt{1+\bm{x}_{e^{*}}^{T}\mathbf{A}^{-1}\bm{x}_{e^{*}}}}$.
\State Compute and save $z_{i}=\bm{v}^{T}\bm{x}_{i}$ for all $i\in \mathcal{N}$. \label{line:zi2}
\State Compute and save $d_{e}=d_{e}-(z_{i}-z_{j})^{2}$ for all $e\in \mathcal{C}\setminus\mathcal{S}$.
\State $\mathbf{A}^{-1}=\mathbf{A}^{-1}-\bm{v}\bm{v}^{T}$.
\EndProcedure
\end{algorithmic}
\end{scriptsize}
\end{algorithm}

In both previous algorithms, $d_{e}$ is computed from scratch, not taking advantage of the previous iteration's computation. Let $d_{e}$, $d_e'$ be the values of the (proxy) marginal gain for $e$ at iterations $k$ and $k+1$, respectively. By the Sherman Morrison formula:
\begin{small}
\begin{equation}
d_{e}^{'}=\textstyle d_{e}-\bm{x}_{e}^{T}\frac{\mathbf{A}^{-1}\bm{x}_{e^{*}}\bm{x}_{e^{*}}^{T}\mathbf{A}^{-1}}{1+\bm{x}_{e^{*}}^{T}\mathbf{A}^{-1}\bm{x}_{e^{*}}}\bm{x}_{e}
=d_{e}-(\bm{x}_{e}^{T}\bm{v})^{2},\label{eq:adaptde}
\end{equation}
\end{small}
 where 
$\textstyle\bm{v}\equiv \frac{\mathbf{A}^{-1}\bm{x}_{e^{*}}}{\sqrt{1+\bm{x}_{e^{*}}^{T}\mathbf{A}^{-1}\bm{x}_{e^{*}}}}.$ 
Exploiting the  pairwise structure $\bm{x}_{e}=\bm{x}_{i}-\bm{x}_{j}$, we get:
\begin{small}
\begin{align}\label{eq:scalar}
\textstyle d_{e}^{'}=d_{e}-(z_{i}-z_{j})^{2},~~ e=(i,j)\in \Omega\setminus \mathcal{S},
\end{align}
\end{small}
where $z_{i}\equiv\bm{v}^{T}\bm{x}_{i}$.
This gives rise to our final greedy implementation, summarized in Alg.~\ref{alg:sg}. 
In  \textsc{PreProcessing}, we factorize matrix $\mathbf{A}^{-1}$ into $\mathbf{A}^{-1}=\mathbf{U}^{T}\mathbf{U}$ and calculate $d_{e}$ for all $e\in\mathcal{C}$ via Eq.~\eqref{eq:factor}. 
In \textsc{UpdateS}, we compute vector $\bm{v}$ 
and scalars $z_{i}=\bm{v}^{T}\bm{x}_{i}$ for all $i\in\mathcal{N}$. Then we update every $d_{e}$ through Eq. \eqref{eq:scalar}, and $\mathbf{A}^{-1}$ via Eq. \eqref{eq:sherman}, using $\bm{v}$ again.

Preprocessing requires $O(d^{2.37})$ time for the matrix inversion and Cholesky factorization, $O(Nd^{2})$ for computing all $\bm{z}_i$, $i\in\mathcal{N}$, and $O(N^{2}d)$ for computing all $d_e$, $e\in \mathcal{C}$.  Computing $\bm{v}$ and updating $\mathbf{A}^{-1}$ via the Sherman Morrison formula have complexity $O(d^{2})$ for one iteration. Computing $\bm{v}^{T}\bm{x}_{i}$, for all $i\in\mathcal{N}$ involves $O(Nd)$ computations, while updating $d_{e}$, for all $e\in \Omega\setminus \mathcal{S}$ via Eq.~\eqref{eq:scalar} requires $O(N^{2})$ scalar computations. Hence, the total complexity is $O(N^{2}(K+d)+N(dK+d^{2})+d^{2}K)$. The $O(N^2(K+d))$ term, due to the computation of $d_e$, $e\in \mathcal{C}$ in preprocessing and at each iteration,  dominates the rest when $N\gg K,d$. The constant in this term thus involves only the time to perform 1 scalar subtraction and 1 scalar multiplication; as such, it remains tractable even for large datasets. 

\section{Accelerating the Lazy Greedy Algorithm}\label{sec:aclazy}

The \emph{lazy greedy} algorithm \cite{minoux1978accelerated,leskovec2007cost,krause2007near} is a well-known variant of the  standard greedy algorithm; it reduces  execution time  by avoiding the computation of all $|\Omega\setminus \mathcal{S}|$ marginal gains $\Delta(e|\mathcal{S})$ at each iteration. This is accomplished via a ``lazy'' evaluation of each inner loop in \textsc{FindMax} in Alg.~\ref{alg:greedy}; though no bounds exist on the worst-case amortized complexity of lazy greedy, it performs quite well in practice \cite{golovin2011adaptive,changshui2011fast}. 

We  employ the same optimizations we describe in Sec.~\ref{sec:aca} to also accelerate the lazy greedy algorithm. 
Each of the three accelerations we mentioned in the previous section yield corresponding ``lazy'' versions, namely \emph{Na\"ive Lazy Greedy}, \emph{Factorization Lazy Greedy}, and \emph{Scalar Lazy Greedy}, respectively. A full description of  these three versions \techrep{can be found in the extended version of this paper \cite{fullversion}}{is in Appendix \ref{sec:AppendixA} }.

In the latter two cases (Factorization and Scalar Lazy Greedy), an additional form of accelaration can be used. Due to lazy evaluation, not all quantities such as, e.g., $z_i$ (in Line \ref{line:zi1} of Alg.~\ref{alg:fga}  and Line \ref{line:zi2} of Alg.~\ref{alg:sg}) are used throughout an iteration. Such quantities can either be pre-computed at each iteration, or computed on the spot, as needed. Though the latter appears to be a faster approach, in practice, it is not always the case:  pre-computation can be faster, as matrix-vector multiplication is more efficient than  for-loops in many languages.  As discussed in Sec.~\ref{sec:evaluation}, we implement both variants in  python, and refer to them as \emph{with-precomputation} and \emph{with-memoization}, respectively.

\section{Evaluation}\label{sec:evaluation}

\begin{figure*}[!t]
\begin{minipage}[b]{0.70\textwidth}
\scriptsize
\vspace{-1em}
\centering
\begin{tabular}{|c||c|c|c|c|c|c|c|c|c|} 
\hline
Dataset & $N$  & $d$ &$|\mathcal{A}|$& $|\mathcal{C}|$ & $|\mathcal{C}_{\texttt{trn}}|$  & $T_{n}(s)$& $A_{\texttt{ab}}$&$A_{\texttt{ac}}$  & $\lambda$\\
 \hline
ROP & 100 & 156  & 30 & 4950  & 1770&6.30  &0.938& 0.858& 0.0001\\
\hline
Sushi & 100 & 20  & 15 & 4821  & 1560--1762 &\!0.878\!&0.932& 0.682&0.0001\\
\hline
Netflix & \!833--1198\! & 30  & 20 & \!180K--540K\! & \!160K--450K\! &88.4&0.811& 0.871&0.0001
\\
\hline
\!CAMRa\! & \!896--3300\! & 10  & 20 & \!400K--5M\!  & \!400K-5M\!  &459 &0.77 &0.79 &0.0001\\
\hline
SIFT & 3000 & 128  & 30 & 4.5M  & 4.5M &12K& N/A & N/A &0.001\\
 \hline
ROP5K & 3000 & 143  & 30 & 4.5M  & 4.5M  &14K & N/A& N/A&0.0001\\
 \hline
MSLR &325-996 & 134  & 30 & \!52K--500K\!  & \!52K--500K\!  &252 &N/A & N/A&0.0001\\
 \hline

\end{tabular}\\
\medskip
(a) Dataset Summary
\end{minipage}
\begin{minipage}[b]{0.35\textwidth}
\begin{center}
\scriptsize
\includegraphics[width=0.8\columnwidth]{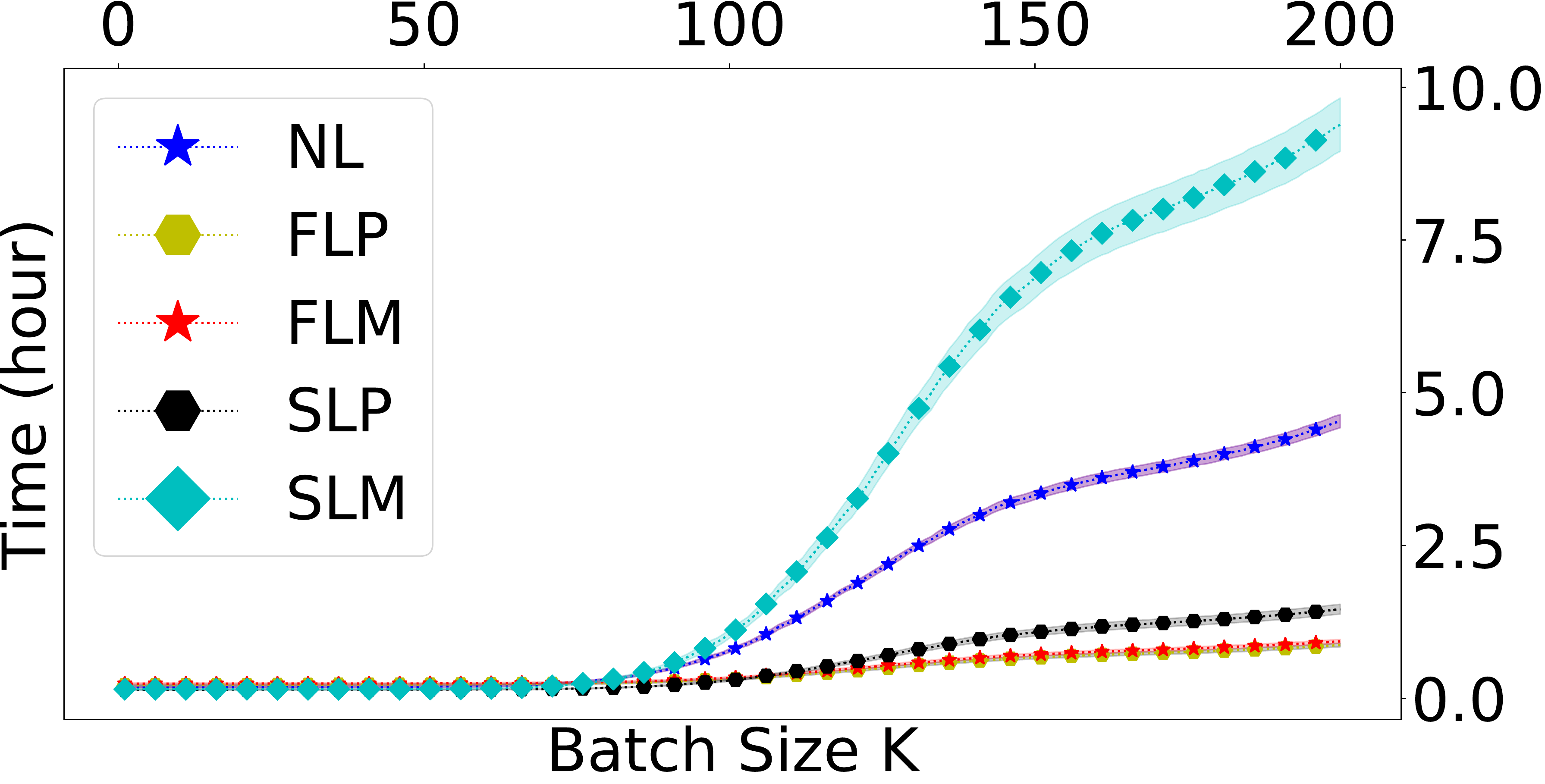}\\
(b) Scalability
\end{center}
\end{minipage}
\vskip -0.12in
\caption{ \emph{(a)} Summary of real datasets. 
Columns $N$, $d$, $|\mathcal{A}|$, and $|\mathcal{C}|$ indicate the number of samples, the dimension, the number of absolute labels, and the number of comparisons, respectively. Column 
$|\mathcal{C}_{\texttt{trn}}|$ is size of train comparison set, while
$T_{\texttt{n}}$ is the execution time under NG for $\mathcal{S}\subseteq \mathcal{C}_{\texttt{trn}},|\mathcal{S}|=200$. Columns $A_{\texttt{ab}},A_{\texttt{ac}}$, indicate the test set AUC of absolute and comparison labels, respectively, when $\mathcal{S}=\mathcal{C}_{\texttt{trn}}$; we report these only for datasets for which we have comparisons. 
Finally, $\lambda$ is the positive value in Eq.~\eqref{eq:dobj}. For Netflix  and Camra, we indicate value ranges across 150 users as appropriate; for MSLR, we report ranges for 150 queries. \emph{(b)} Scalability of Lazy Greedy Algorithm. Time execution result for large synthetic dataset with  $N=15000$, $d=400$, and $|\mathcal{C}|=1.125\times 10^{8}$. }\label{table:dataset}\label{fig:Scalability}
\vskip -0.2in
\end{figure*}



We use  synthetic and real datasets to evaluate the performance of different greedy and lazy greedy algorithms.\footnote{\vspace*{-3mm}\scalebox{1}{
  Our code is publicly available at: \url{https://github.com/neu-spiral/AcceleratedExperimentalDesign}}} We evaluate these algorithms both  in terms of execution time and classification performance w.r.t.~accuracy of predictions, after labels are collected.

\subsection{Evaluation Setup.}
We begin by describing our evaluation setup.

\noindent\textbf{Datasets.} 
In our synthetic dataset,  the absolute feature vectors $\bm{x}_{i}\in\mathbb{R}^d$, $i\in \mathcal{N}$, are sampled from a Gaussian distribution $\normal(\bm{0},\sigma_{x}I_{d})$ with feature dimension $d$ ranging from $20$ to $400$ and dataset size $N$ ranging from $500$ to $15000$.  We also sample a parameter vector $\widetilde{\bm{\beta}}$ from Gaussian distribution $\normal(\bm{0},\sigma_{\bm{\beta}}I_{d})$.  We generate absolute labels $y_i$, $i\in \mathcal{N}$, using Eq.~\eqref{eq1} with $\bm{\beta}=\widetilde{\bm{\beta}}/C_{a}$, where $C_{a}$ is a positive scalar. 
Finally, we generate $|\mathcal{C}|$ comparison labels via Eq.~\eqref{eq2}, with $\bm{\beta}=\widetilde{\bm{\beta}}$. Parameter $C_a$ allows us to control the relative noise ratio between absolute and comparison labels; we set it to $C_a=1.2$ in our experiments.

We also use seven real-life datasets, summarized in  Fig.~\ref{table:dataset}(a). The first four (ROP, Sushi, Netflix, Camra) contain comparison labels; the remaining (ROP5K, SIFT,  and Microsoft URL)  do not, and are used only for measuring the execution time of our algorithms. A detailed description of all  datasets is in \techrep{\cite{fullversion}}{Appendix \ref{sec:AppendixB}}.

\noindent\textbf{Algorithms.}
We implement eight greedy algorithms:  Na\"ive Greedy (NG), Factorization Greedy (FG),  Scalar Greedy (SG),  Na\"ive Lazy Greedy (NL),  Factorization Lazy Greedy with Pre-Computation (FLP), Factorization Lazy Greedy with Memoization (FLM),  Scalar Lazy Greedy with Pre-Computation (SLP), and Scalar Lazy Greedy with Memoization (SLM). 
In each dataset, we set $\lambda$ in \eqref{eq:dobj} to about $10^{-5}$ the average norm of feature vectors (see Fig~\ref{table:dataset}(a)).     

We also implement  the greedy algorithm with Mutual Information (Mut), Fisher Information (Fisher), and Entropy (Ent) objectives, as described in \cite{guo} (also reviewed in  \techrep{\cite{fullversion}}{Appendix \ref{sec:b2}}). Finally, we implement a Random (Ran) baseline method, in which the set $\mathcal{S}$ is selected uniformly at random from $\mathcal{C}_{trn}$.


\noindent\textbf{Experiment Setup.}
In each experiment, we partition the dataset $\mathcal{N}$ into three datasets: a training set $\mathcal{N}_{\mathtt{trn}}$, a test set $\mathcal{N}_{\mathtt{tst}}$, and a validation set $\mathcal{N}_{\mathtt{val}}$.
Wherever available, we denote by $\mathcal{C}_{\mathtt{trn}}\subset \mathcal{C}$  the corresponding comparison set restricted to pairs of objects in $\mathcal{N}_{\mathtt{trn}}$. We  select a random subset $\mathcal{A}$ from $\mathcal{N}_{\mathtt{trn}}$ whose absolute labels $y_{i}$, $i\in \mathcal{A}$ are presumed revealed to the experimenter. Then we use our greedy algorithms to select $\mathcal{S} \subset \mathcal{C}_{\mathtt{trn}}, |\mathcal{S}|=K$.
We record the running time $t_{K}$ of each algorithm for different values of $K\in \mathbb{N}$ executed on the training set. For synthetic data, we repeat each experiment 150 times, each time with a different randomly generated dataset; we report average $t_k$ values, as well as standard deviations. For real datasets including absolute labels (ROP, Sushi, Netflix, Camra), we also repeat experiments 150 times, each time with a different randomly selected set $\mathcal{A}$.

For both synthetic and real datasets for which we have comparison labels (ROP, Sushi, Netflix, Camra), 
we collect the $K$ comparison labels from $\mathcal{S}$ and train a model $\bm{\beta}\in\mathbb{R}^d$ using the labels in $\mathcal{A}$ and $\mathcal{S}$ via MAP estimation \eqref{eq4}, and predict both comparison and absolute labels in the test set. In doing so, we select the parameter $\lambda$ in \eqref{eq4} as the value that maximizes AUC on the validation set. Especially, for ROP, we measure the performance w.r.t. the reference standard diagnosis (RSD) label prediction rather than absolute labels,  even though the model is trained on (noisier) absolute labels. 
For each dataset, we perform cross validation, repeating the partition to training and test datasets and keeping the validation set fixed. To produce confidence intervals, each 4-fold cross validation is repeated 150 times, i.e., over 150 different random data shuffles (for the Netflix dataset, the experiment is  executed for 150 users).  

\begin{figure}[!t]
\begin{center}
\centerline{\includegraphics[width=0.8\columnwidth,height=0.5\columnwidth]{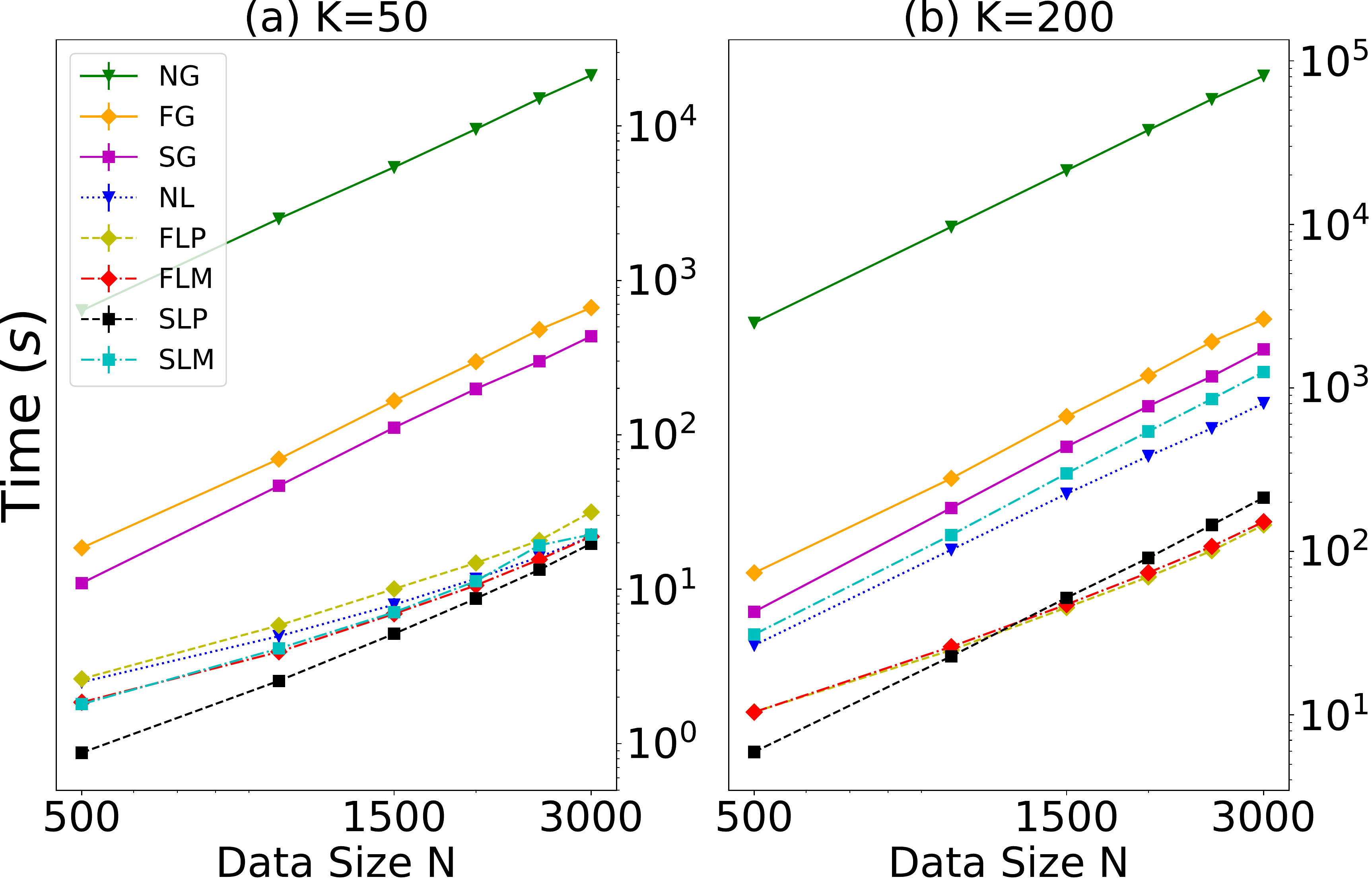}}
\caption{Average execution time for synthetic data under different sample sizes $N$, with feature dimension $d=400$ and $K$ set to $K=50$ in subfigure (a) and $K=200$ in subfigure (b).}
\label{fig:tvn}
\end{center}
\vskip -0.4in
\end{figure}

\begin{figure}[!t]
\begin{center}
\centerline{\includegraphics[width=0.8\columnwidth,height=0.5\columnwidth]{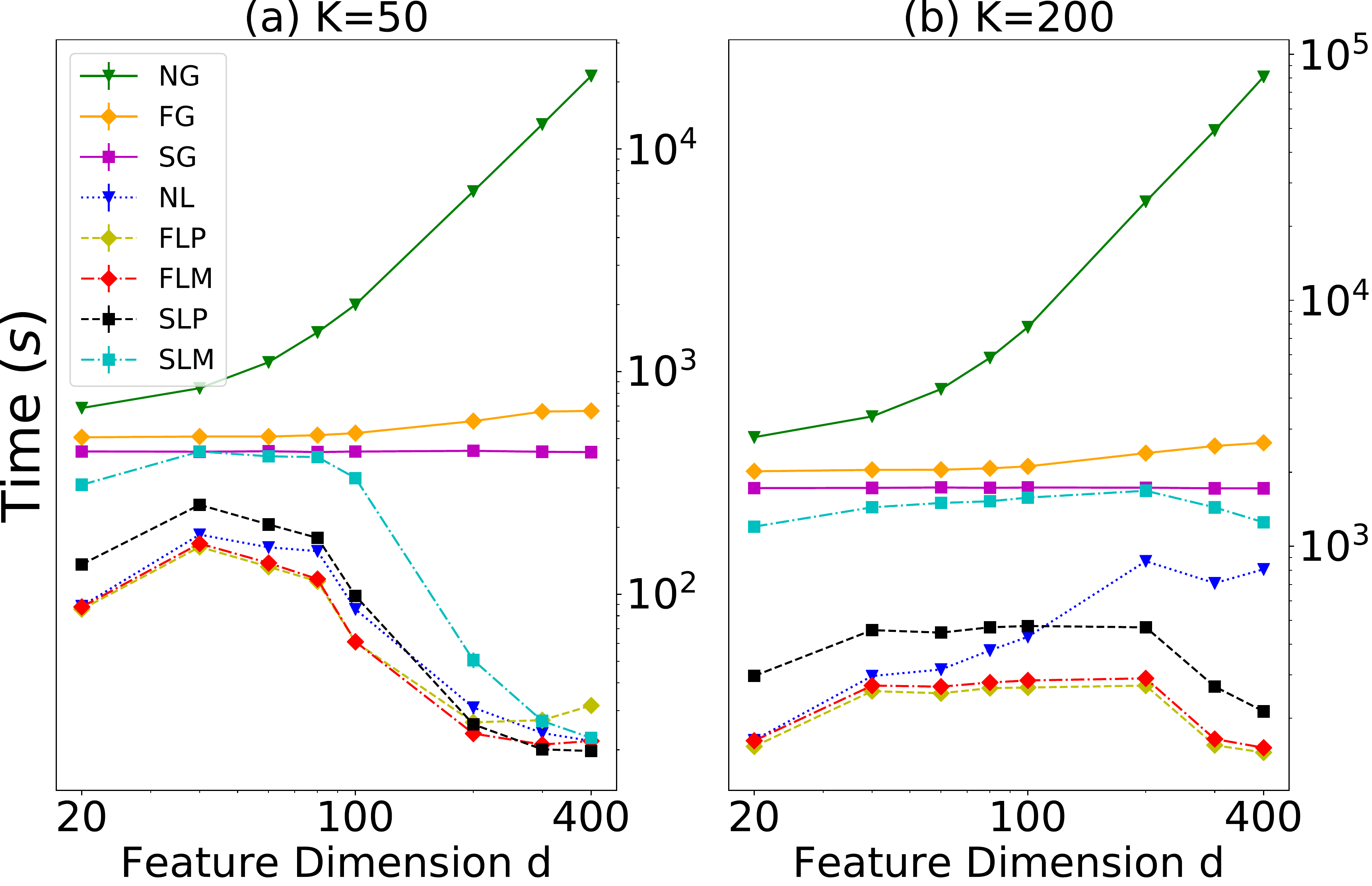}}
\caption{Average execution time  for  synthetic data under different feature dimensions $d$, with sample size $N=3000$ and $K$ set to $K=50$ in subfigure (a) and $K=200$ in subfigure (b).}
\label{fig:tvd}
\end{center}
\vskip -0.3in
\end{figure}

\begin{figure}[!t]
\begin{center}
\centerline{\includegraphics[width=1.0\columnwidth,height=0.50\columnwidth]{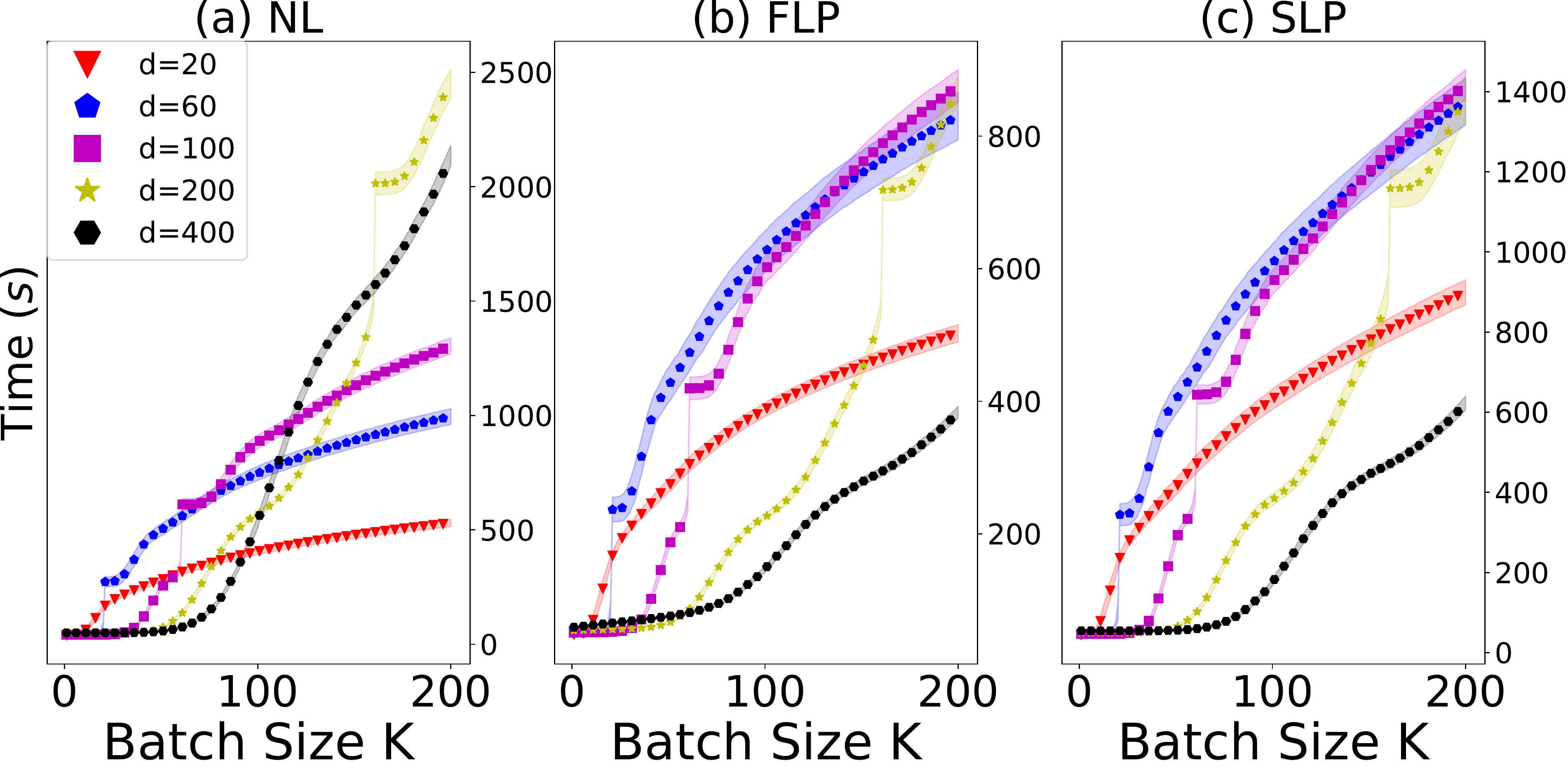}}

\caption{Average execution time as a function of $K$ on synthetic data with $N=5000$ for three lazy greedy algorithms.}

\label{fig:tvk}
\end{center}
\vskip -0.4in
\end{figure}
\begin{figure}[!t]
\begin{centering}
\centerline{\includegraphics[width=\columnwidth]{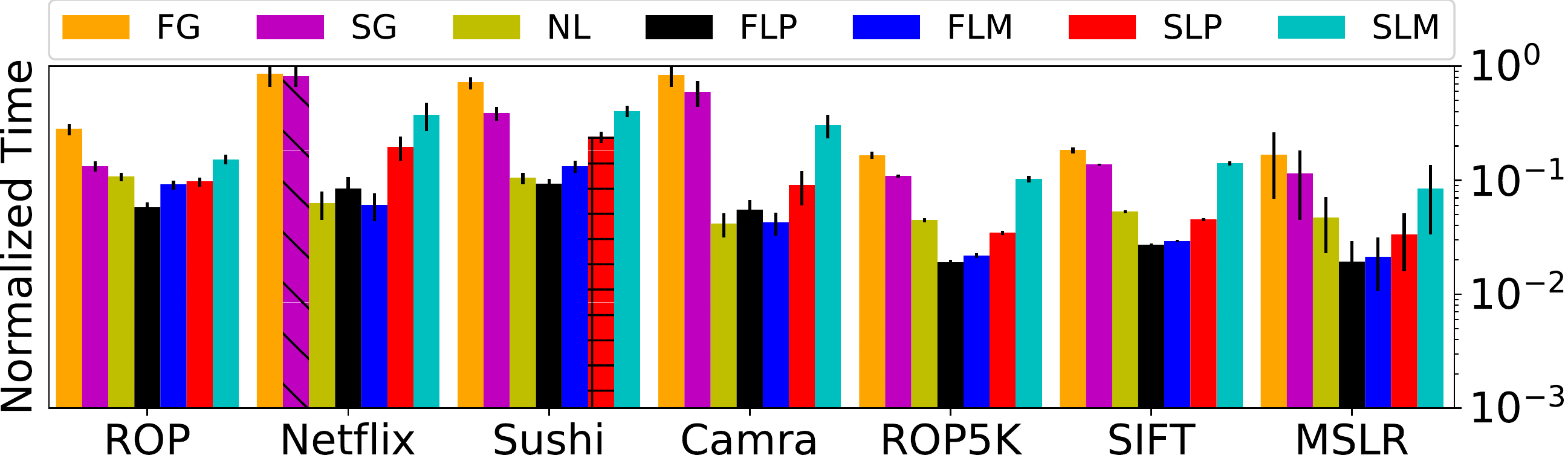}}

\caption{Execution time for different datasets and different algorithms, normalized by NG execution time (see col.~$T_n(s)$ in Fig.~\ref{table:dataset}(a)).}
\label{fig:NormalTime}
\end{centering}
\vskip -0.2in
\end{figure}



\subsection{Execution Time Performance.}
We first study the execution time in terms of $N$, $d$, and $K$.


\noindent\textbf{Dependence on $N$.} In Fig.~\ref{fig:tvn} we plot the running time as a function of the data size $N$ for  synthetic datasets. The quadratic--$O(N^2)$--scaling of all algorithms is clearly evident, although the actual execution time varies drastically between different algorithms. Both FG and SG improve upon NG by almost two orders of magnitude. Lazy algorithms improve over NG by as much as 3 orders of magnitude when $K=50$. 
However, scalar lazy greedy with memoization (SLM)  performs similarly to FG and SG when $K=200$, even worse than NL. 

Almost universally, pre-processing versions (FLP and SLP) outperform the corresponding memoized versions of the lazy algorithms (FLM and SLM). This is because pre-computation involves a matrix-vector multiplication: in python's NumPy library this is performed in C language, and is more efficient than the python for-loop inherent in memoization. This negates any benefit of computing only the values needed via memoization. Finally, SLP is the best performer when $K=50$, while FLP outperforms it for large $N$ when $K=200$.
%

\noindent\textbf{Dependence on $d$.} Fig.~\ref{fig:tvd} shows performance over synthetic datasets as a function of dimension $d$. The advantage of FG and SG over the na\"ive algorithm (NG) is clearly evident: the latter grows super-linearly in $d$. In contrast, the effect of $d$ on FG is linear, while on SG it is almost imperceptible.
A striking difference in behavior is observed in the lazy greedy algorithms, that are very sensitive to $d$. Indeed, these algorithms perform \emph{poorly in lower dimensions}, with the gap between performance for low to high dimensions being sometimes close to two orders of magnitude. This is because, for high $d$,  there is are many new dimensions to discover; as a result, almost maximal elements in the heap remain almost maximal in subsequent iterations, leading to early loop terminations. In contrast, in low $d$, maximality changes drastically between iterations, leading to full loop executions. 

\noindent\textbf{Dependence on $K$}. We further explore this phenomenon in Fig.~\ref{fig:tvk}, that shows the dependence of lazy algorithm on $K$.
 We observe a `jump' in execution time, indicating an expensive loop execution that contributes highly to the execution cost. The smaller  $d$ is, the earlier this jump is observed. 

\noindent\textbf{Scalability.} All in all, we observe that  our accelerations, on both standard and lazy greedy algorithms, can significantly reduce the execution time of experimental design. In Fig.~\ref{fig:Scalability}(b), we illustrate this by running the accelerated lazy algorithms for a large synthetic dataset with $N=15000$ and $d=400$, containing more than $10^{8}$ comparison pairs. The running time of Na\"ive Greedy (NG) on this dataset exceeds 10 days. As seen in Fig.~\ref{fig:Scalability},  the running time can be shortened to less than $1$ hour under the Factorization Lazy (FLP) algorithm. We also observe that SLM  performs worse than NL, while SLP outperforms NL,  again due to  the advantage of matrix-vector multiplications over python for-loops. 


\noindent\textbf{Time Performance Evaluation on Real Datasets.}
Experiments on the seven real datasets further corroborate  observations made over synthetic data. 
 Fig.~\ref{fig:NormalTime} shows  the execution time  normalized by the execution time of NG for each dataset (see column $T_n(s)$ in  Fig.~\ref{table:dataset}(a)). 
 All algorithms yield an improvement. Lazy greedy algorithms perform well overall, but for the Sushi, CAMRa, and Netflix datasets, this improvement is diminished due to their low  dimension $d$. 
 The highest improvement in all algorithms compared to NG is observed in the largest of our datasets, ROP5K, where FG and SG yield an improvement of 1 order of magnitude, while SLP performs exceedingly well, leading to an improvement of 2 orders of magnitude over NG. Overall,  FLP consistently improves performance over NL.

\begin{figure}[!t]
\centering
\includegraphics[width=1.0\columnwidth, height=0.4\columnwidth]{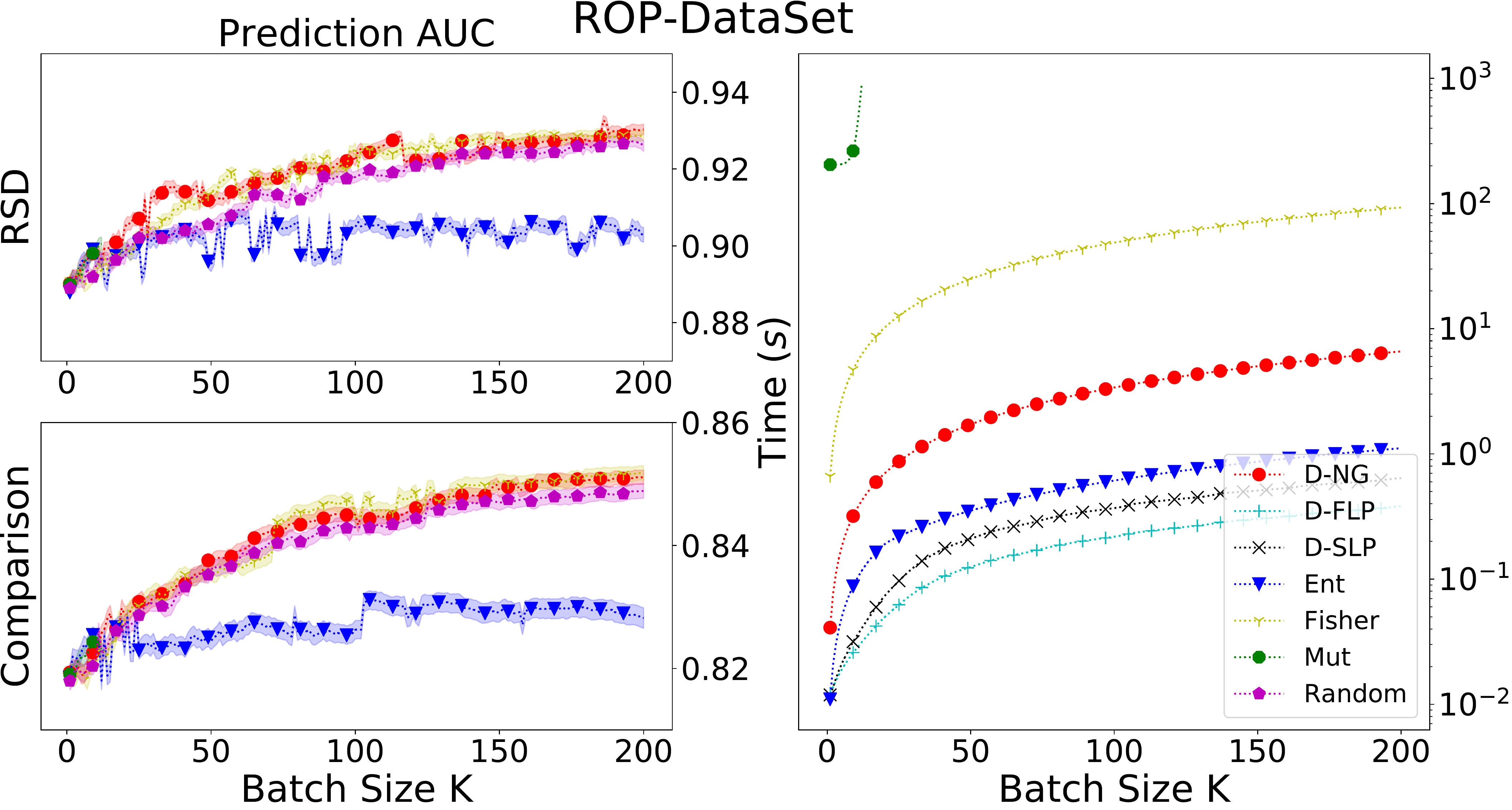}
\caption{Test set AUC and execution time for ROP dataset, when comparisons samples are selected via D-optimal, Fisher, Entropy and Random and Mutual Information. The classifier is trained via MAP \eqref{eq4} on the training set.  The left top figure is the test AUC for RSD label, the left bottom figure is the test AUC for comparison labels. The right figure is the execution time for different algorithms. For the D-optimal objective, we record the execution time for Naive Greedy, Factorization Lazy and Scalar Lazy Greedy Algorithm. We only execute Mutual Information for $K\leq 12$.}
\label{fig:aucrop}
\vskip -0.3in
\end{figure}

\subsection{Prediction Performance}

All 8 algorithms using D-Optimality as an objective produce the same selected set $\mathcal{S}$ (namely, the one determined by the greedy algorithm). We give some intuition of the quality of the model learned via MAP estimation \eqref{eq4} in comparison to competitors. This selection has been known to outperform competitors such as Fisher Information and Entropy objectives~\cite{guo}. For the sake of completeness, we show  in Fig.~\ref{fig:aucrop}
 the prediction quality of the resulting trained model w.r.t AUC of both absolute and comparison labels over the test set,
 on the ROP dataset. Remaining datasets for which we have comparison labels  (Sushi, Netflix, and Camra) are shown in \techrep{\cite{fullversion}}{Appendix \ref{sec:AppendixD}}. In all cases, estimators learned over labels collected by the greedy algorithm significantly outperform random selection. Fisher Information and Mutual Information are sometimes better, but are also exceedingly time consuming, between $10^2-10^3$ times slower than D-optimal NG. Finally, Entropy is fast, but the prediction performance is not as good as under NG. 

We observe similar performance in the remaining datasets, shown in \techrep{\cite{fullversion}}{Appendix \ref{sec:AppendixD}}. For the Netflix and Camra datasets,  we can only scale Mutual Information and Fisher Information to $K\leq 10$, as their complexity is $O(N^2 2^{K})$ and $O(N^{4}K)$, respectively. In the Sushi and Netflix dataset, the best AUC comes from the D-optimal design algorithm. For the ROP dataset, the D-optimal and Fisher information outperform random for all batch sizes. For the Camra dataset, as $d=10$, the D-optimal design algorithm outperforms Random only when the batch size is less than 100. Finally, Entropy  is time efficient but has worse accuracy than all other methods. Our D-optimal Na\"ive Greedy and its variants have good accuracy and are time-efficient; accelarated methods FLP and SLP are even faster than Entropy.

\section{Conclusion}\label{sec:conclusion}
We have shown that experimental design for pairwise comparisons under the D-optimality criterion can be significantly accelerated by exploiting the underlying geometry of pairwise comparisons. Given the prevalence of submodularity in batch active learning objectives, it would be interesting to identify methods through which these results could extend to other objectives of interest. These include objectives that are structurally similar (such as A-optimality or E-optimality \cite{boyd2004convex}), as well as objectives like mutual information, for which even the function value oracle is not tractable.



\begin{footnotesize}
\section{Acknowledgement}
Our work is supported by NIH (R01EY019474, P30EY10572), NSF (SCH-1622542 at MGH; SCH-1622536 and CCF-1750539 at Northeastern; SCH-1622679 at OHSU), and by unrestricted departmental funding from Research to Prevent Blindness (OHSU).
\end{footnotesize}

\techrep{\begin{footnotesize}
\bibliographystyle{yuan}
\bibliography{citation}

\begin{thebibliography}{10}

\bibitem{kalpathy2016plus}
J.~K-Cramer, J.~P. Campbell, D.~Erdogmus, et~al.
\newblock {\em Plus disease in retinopathy of prematurity: improving diagnosis
  by ranking disease severity and using quantitative image analysis}.
\newblock Ophthalmology, 2016.

\bibitem{sculley2010combined}
D.~Sculley.
\newblock {\em Combined regression and ranking}.
\newblock In KDD, 2010.

\bibitem{desarkar2010aggregating}
M.~S. Desarkar, S.~Sarkar, and P.~Mitra.
\newblock {\em Aggregating preference graphs for collaborative rating
  prediction}.
\newblock In Recsys, 2010.

\bibitem{desarkar2012preference}
M.~S. Desarkar, R.~Saxena, and S.~Sarkar.
\newblock {\em Preference relation based matrix factorization for recommender
  systems}.
\newblock In UMAP, 2012.

\bibitem{guo}
Y.~Guo, P.~Tian, J.~Kalpathy-Cramer, S.~Ostmo, J.~P. Campbell, M.~F. Chiang,
  D.~Erdogmus, J.~Dy, and S.~Ioannidis.
\newblock {\em Experimental Design Under the {Bradley-Terry} Model}.
\newblock In IJCAI, 2018.

\bibitem{stewart2005absolute}
N.~Stewart, G.~DA Brown, and N.~Chater.
\newblock {\em Absolute identification by relative judgment.}
\newblock Psychological Review, 2005.

\bibitem{brun2010towards}
A.~Brun, A.~Hamad, O.~Buffet, and A.~Boyer.
\newblock {\em Towards preference relations in recommender systems}.
\newblock In ECML/PKDD, 2010.

\bibitem{zheng2009mining}
Y.~Zheng, L.~Zhang, X.~Xie, and Wei-Ying Ma.
\newblock {\em Mining interesting locations and travel sequences from GPS
  trajectories}.
\newblock In WWW. ACM, 2009.

\bibitem{koren2011ordrec}
Y.~Koren and J.~Sill.
\newblock {\em OrdRec: an ordinal model for predicting personalized item rating
  distributions}.
\newblock In Recsys, 2011.

\bibitem{schultz2004learning}
M.~Schultz and T.~Joachims.
\newblock {\em Learning a distance metric from relative comparisons}.
\newblock In NIPS, 2004.

\bibitem{boyd2004convex}
S.~Boyd and L.~Vandenberghe.
\newblock {\em Convex optimization}.
\newblock Cambridge university press, 2004.

\bibitem{pukelsheim1993optimal}
F.~Pukelsheim.
\newblock {\em Optimal design of experiments}.
\newblock SIAM, 1993.

\bibitem{nemhauser1978}
G.~L. Nemhauser, L.~A. Wolsey, and M.~L Fisher.
\newblock {\em An analysis of approximations for maximizing submodular set
  functions}.
\newblock Mathematical Programming, 1978.

\bibitem{van1996matrix}
G.~H. Golub and C.~F. Van~Loan.
\newblock {\em Matrix computations}.
\newblock JHU Press, 2012.

\bibitem{sherman1950adjustment}
J.~Sherman and W.~J. Morrison.
\newblock {\em Adjustment of an inverse matrix corresponding to a change in one
  element of a given matrix}.
\newblock Ann. Math. Stat, 1950.

\bibitem{minoux1978accelerated}
M.~Minoux.
\newblock {\em Accelerated greedy algorithms for maximizing submodular set
  functions}.
\newblock In Optimization techniques. 1978.

\bibitem{mirzasoleiman2015lazier}
B.~Mirzasoleiman, A.~Badanidiyuru, A.~Karbasi, J.~Vondr{\'a}k, and A.~Krause.
\newblock {\em Lazier Than Lazy Greedy.}
\newblock In AAAI, 2015.

\bibitem{lin2011class}
H.~Lin and J.~Bilmes.
\newblock {\em A class of submodular functions for document summarization}.
\newblock In HLT, 2011.

\bibitem{mirzasoleiman2013distributed}
B.~Mirzasoleiman, A.~Karbasi, R.~Sarkar, and A.~Krause.
\newblock {\em Distributed submodular maximization: Identifying representative
  elements in massive data}.
\newblock In NIPS, 2013.

\bibitem{chen2015fusing}
L.~Chen, P.~Zhang, and B.~Li.
\newblock {\em Fusing pointwise and pairwise labels for supporting
  user-adaptive image retrieval}.
\newblock In ICMR, pp 67--74, 2015.

\bibitem{takamura2015estimating}
H.~Takamura and J.~Tsujii.
\newblock {\em Estimating numerical attributes by bringing together fragmentary
  clues}.
\newblock In HLT, 2015.

\bibitem{wang2016ppp}
Y.~Wang, S.~Wang, J.~Tang, H.~Liu, and B.~Li.
\newblock {\em {PPP}: Joint pointwise and pairwise image label prediction}.
\newblock In CVPR, 2016.

\bibitem{liepe2013maximizing}
J.~Liepe, S.~Filippi, M.~Komorowski, and M.~P. Stumpf.
\newblock {\em Maximizing the information content of experiments in systems
  biology}.
\newblock PLOS Comput. Biol, 2013.

\bibitem{cavagnaro2010adaptive}
D.~R. Cavagnaro, J.~I. Myung, M.~A. Pitt, and J.~V. Kujala.
\newblock {\em Adaptive design optimization: A mutual information-based
  approach to model discrimination in cognitive science}.
\newblock Neural computation, 2010.

\bibitem{krause2012near}
A.~Krause and C.~E. Guestrin.
\newblock {\em Near-optimal nonmyopic value of information in graphical
  models}.
\newblock arXiv preprint arXiv:1207.1394, 2012.

\bibitem{busetto2013near}
A.~G. Busetto, A.~Hauser, G.~Krummenacher, M.~Sunn{\aa}ker, S.~Dimopoulos,
  C.~S. Ong, J{\"o}. Stelling, and J.~M. Buhmann.
\newblock {\em Near-optimal experimental design for model selection in systems
  biology}.
\newblock Bioinformatics, pp 2625--2632, 2013.

\bibitem{krause2014submodular}
A.~Krause and D.~Golovin.
\newblock {\em Submodular function maximization.}, 2014.

\bibitem{golovin2011adaptive}
D.~Golovin and A.~Krause.
\newblock {\em Adaptive submodularity: Theory and applications in active
  learning and stochastic optimization}.
\newblock JAIR, 2011.

\bibitem{active}
K.~G. Jamieson and R.~Nowak.
\newblock {\em Active ranking using pairwise comparisons}.
\newblock In NIPS, 2011.

\bibitem{grasshoff2008optimal}
U.~Gra{\ss}hoff and R.~Schwabe.
\newblock {\em Optimal design for the Bradley--Terry paired comparison model}.
\newblock Statistical Methods and Applications, 2008.

\bibitem{glickman2005adaptive}
M.~E. Glickman and S.~T. Jensen.
\newblock {\em Adaptive paired comparison design}.
\newblock Journal of statistical planning and inference, pp 279--293, 2005.

\bibitem{bradley1952rank}
R.~A. Bradley and M.~E. Terry.
\newblock {\em Rank analysis of incomplete block designs: I. The method of
  paired comparisons}.
\newblock Biometrika, 1952.

\bibitem{he2010laplacian}
X.~He.
\newblock {\em Laplacian regularized D-optimal design for active learning and
  its application to image retrieval}.
\newblock IEEE Transactions on Image Processing, 2010.

\bibitem{harville1997matrix}
D.~A. Harville.
\newblock {\em Matrix algebra from a statistician's perspective}.
\newblock Springer, 1997.

\bibitem{press2007numerical}
W.~H. Press.
\newblock {\em Numerical recipes 3rd edition: The art of scientific computing}.
\newblock Cambridge university press, 2007.

\bibitem{leskovec2007cost}
J.~Leskovec, A.~Krause, C.~Guestrin, C.~Faloutsos, J.~VanBriesen, and
  N.~Glance.
\newblock {\em Cost-effective outbreak detection in networks}.
\newblock In KDD, 2007.

\bibitem{krause2007near}
A.~Krause and C.~Guestrin.
\newblock {\em Near-optimal observation selection using submodular functions}.
\newblock In AAAI, 2007.

\bibitem{changshui2011fast}
Z.~Changshui, H.~Guangdong, and W.~Jun.
\newblock {\em A fast algorithm based on the submodular property for
  optimization of wind turbine positioning}.
\newblock Renewable Energy, 2011.

\bibitem{liu2011entropy}
M.~Liu, O.~Tuzel, S.~Ramalingam, and R.~Chellappa.
\newblock {\em Entropy rate superpixel segmentation}.
\newblock In CVPR, 2011.

\bibitem{gomez2012inferring}
M.~Gomez-Rodriguez, J.~Leskovec, and A.~Krause.
\newblock {\em Inferring networks of diffusion and influence}.
\newblock TKDD, 2012.

\bibitem{calinescu2011maximizing}
G.~Calinescu, C.~Chekuri, M.~P{\'a}l, and J.~Vondr{\'a}k.
\newblock {\em Maximizing a monotone submodular function subject to a matroid
  constraint}.
\newblock SIAM Journal on Computing, 2011.

\bibitem{krause2008near}
A.~Krause, A.~Singh, and C.~Guestrin.
\newblock {\em Near-optimal sensor placements in Gaussian processes: Theory,
  efficient algorithms and empirical studies}.
\newblock JMLR, 2008.

\bibitem{sun2015submodboxes}
Q.~Sun and D.~Batra.
\newblock {\em Submodboxes: Near-optimal search for a set of diverse object
  proposals}.
\newblock In NIPS, 2015.

\bibitem{brown2018fully}
J.~M. Brown, J.~P. Campbell, A.~Beers, K.~Chang, K.~Donohue, S.~Ostmo, RV.~P.
  Chan, J.~Dy, D.~Erdogmus, S.~Ioannidis, et~al.
\newblock {\em Fully automated disease severity assessment and treatment
  monitoring in retinopathy of prematurity using deep learning}.
\newblock In Medical Imaging, 2018.

\bibitem{kamishima2009trbagg}
T.~Kamishima, M.~Hamasaki, and S.~Akaho.
\newblock {\em A simple transfer learning method and its application to
  personalization in collaborative tagging}.
\newblock In ICDM, 2009.

\bibitem{elo1978rating}
A.~E. Elo.
\newblock {\em The rating of chessplayers past and present}.
\newblock Arco Pub, 1978.

\bibitem{koren2009matrix}
Y.~Koren, R.~Bell, and C.~Volinsky.
\newblock {\em Matrix factorization techniques for recommender systems}.
\newblock Computer, 2009.

\bibitem{DBLP:journals/corr/QinL13}
Tao Qin and Tie{-}Yan Liu.
\newblock {\em Introducing {LETOR} 4.0 Datasets}.
\newblock CoRR, abs/1306.2597, 2013.

\bibitem{Dua:2017}
D.~Dheeru and E.~Karra~Taniskidou.
\newblock {\em {UCI} Machine Learning Repository}, 2017.

\bibitem{bento2011identifying}
J.~Bento, N.~Fawaz, A.~Montanari, and S.~Ioannidis.
\newblock {\em Identifying users from their rating patterns}.
\newblock In CAMRa, 2011.

\end{thebibliography}
\end{footnotesize}}{}

\techrep{}{%

\appendix


\newpage
\section{Accelerating the Lazy Greedy Algorithm}\label{sec:AppendixA}

\begin{algorithm}[!h]
\begin{scriptsize}
\caption{Lazy Greedy Algorithm} \label{alg:lazygreedy}
Lazy Greedy Algorithm, as described in Sec.~\ref{sec:aclazy}. The main \textsc{Greedy} procedure, as well as \textsc{UpdateS}, are the same as in Alg.~\ref{alg:greedy}. Tuples are ordered lexicographically.
\begin{algorithmic}[1]
\Procedure {PreProcessing}{$\mathbf{X}$}
\State For all $e\in \Omega$, insert tuples $(\Delta(e|\emptyset),e)$ into heap $\mathbf{H}$; Set $\mathcal{S} = \emptyset$.
\EndProcedure
\end{algorithmic}
\begin{algorithmic}[1]
\Procedure {FindMax}{$\mathcal{S}$}
\While{$\mathbf{True}$}
\State $(\Delta_{\texttt{old}},e)=$$\mathbf{H}$.\textsc{PopMax}()
\State Set $\Delta=\textsc{UpdateMarginal}(e,S)$
\State $(\Delta',e') = \mathbf{H}.$\textsc{Max}()
\If{$\Delta \geq \Delta'$}
\Return e
\Else
\State $\mathbf{H}$.\textsc{Insert}$\big( (\Delta,e )\big)$
\EndIf
\EndWhile 
\EndProcedure
\end{algorithmic}
\begin{algorithmic}[1]
\Procedure{UpdateMarginal}{$e,S$}
\State \Return $\Delta(e|\mathcal{ S})$
\EndProcedure
\end{algorithmic}
\end{scriptsize}
\end{algorithm}

We can employ the same optimizations we describe in Sec.~\ref{sec:aca} to also accelerate the lazy greedy algorithm. Before describing these optimizations, we briefly review the algorithm below.
Intuitively, the lazy greedy algorithm exploits the following fact: by submodularity, for any sets $\mathcal{S'}\subseteq \mathcal{S}\subseteq \Omega$ and any $e\in \Omega$ we have that
$\Delta(e|\mathcal{S})\leq \Delta(e|\mathcal{S}').$
Every iteration of the greedy algorithm produces a set $\mathcal{S}$ that is a superset of the selected sets $\mathcal{S}'$ at all previous iterations. As a result,   marginal gains $\Delta(e|\mathcal{S}')$ at \emph{any} previous iteration serve as \emph{upper bounds} on  marginal gains $\Delta(e|\mathcal{S})$ in the current iteration. Thus, to discover $e^*$, it suffices to find an element whose current marginal $\Delta(e|\mathcal{S})$ exceeds all past marginal gains: if such an $e$ is found, the loop can terminate early. 

The above observation leads to the lazy greedy algorithm in Alg.~\ref{alg:lazygreedy}. Past marginal gains are stored in a heap\footnote{Recall that a heap is a data structure that supports three operations: $\textsc{Insert}()$, that adds an element from an ordered set; $\textsc{PopMax()}$, that removes and returns the maximum element in the heap; and $\textsc{Max}()$, that returns the maximum element (without removing it). For a heap of size $n$, these operations have  complexity $O(\log n)$,  $O(\log n)$, and $O(1)$, respectively.} \cite{liu2011entropy}. To find $e^*$, the algorithm pops the maximal element $e$ from the heap, and computes its present marginal gain $\Delta(e|\mathcal{S})$.  If this exceeds the gain of the (next) maximal element in the heap, then $e$ is $e^*$: this is because all values in the heap are upper bounds on the true marginal gains. The loop can thus return $e$ and terminate early. Otherwise, $e$ is placed back in the heap with its (updated) $\Delta(e|S)$, and the process repeats. 

Under the value oracle model, the  complexity of the lazy greedy algorithm is in fact worse than the standard greedy algorithm: the worst-case cost of an iteration is $O(|\Omega|\log|\Omega|)$, due to heap operations. Though no amortized complexity results are known, in practice loops often terminate early; this leads to a significant computational improvement in a broad array of problems \cite{gomez2012inferring,calinescu2011maximizing,krause2008near,sun2015submodboxes} and motivates us to apply our accelerations to lazy greedy as well.
We now describe how the accelerations we presented in  Sec.~\ref{sec:aca} can be incorporated in the lazy greedy algorithm. 

\begin{algorithm}[!h]
\begin{scriptsize}
\caption{Na\"ive Lazy Greedy Algorithm} \label{alg:naiveL}
Na\"ive Lazy Algorithm, as described in Sec.~\ref{sec:aclazy}. The main \textsc{Greedy} procedure, as well as \textsc{UpdateS} and \textsc{FindMax} are the same as in Alg.~\ref{alg:lazygreedy}.
\begin{algorithmic}[1]
\Procedure {PreProcessing}{$\mathbf{X}$}
\State Compute
	$\mathbf{A}_{0}^{-1}=(\lambda I_{d}+\underset{i\in\mathcal{A}}{\sum}\bm{x}_{i}\bm{x}_{i}^{T})^{-1}$; Set $\mathbf{A}^{-1}=\mathbf{A}_0^{-1}$
\State For all $e\in \Omega$, insert tuples $(\bm{x}_{e}^{T}\mathbf{A}^{-1}\bm{x}_{e},e)$ into heap $\mathbf{H}$;  Set $\mathcal{S} = \emptyset$.
\EndProcedure
\end{algorithmic}
\begin{algorithmic}[1]
\Procedure{UpdateMarginal}{$e,S$}
\State \Return $\bm{x}_{e}^{T}\mathbf{A}^{-1}\bm{x}_{e}$
\EndProcedure
\end{algorithmic}
\end{scriptsize}
\end{algorithm}

\noindent\textbf{Na\"ive Lazy Greedy.} To begin with, we can exploit the same simple improvements we described in Sec.~\ref{subsec:naive}: rather than storing the marginal gains $\Delta(e|\mathcal{S})$, the--simpler to compute--quantities $d_e$, given by \eqref{eq:de}, can be stored in the heap instead. Matrix $\mathbf{A}^{-1}$, can again be updated  via the Sherman-Morisson formula \eqref{eq:sherman}. Both are straightforward to implement; see Alg.~\ref{alg:naiveL} for pseudocode.

\begin{algorithm}[!t]
\begin{scriptsize}
\caption{Factorization Lazy Greedy Algorithm} \label{alg:factorL}
Factorization Lazy Algorithm, as described in Sec.~\ref{sec:aclazy}. The main \textsc{Greedy} procedure, as well as \textsc{FindMax}, are the same as in Alg.~\ref{alg:lazygreedy}.
\begin{algorithmic}[1]
\Procedure {PreProcessing}{$\mathbf{X}$}
\State Compute
	$\mathbf{A}_{0}^{-1}=(\lambda I_{d}+\underset{i\in\mathcal{A}}{\sum}\bm{x}_{i}\bm{x}_{i}^{T})^{-1}$; Set $\mathbf{A}^{-1}=\mathbf{A}_0^{-1}$; Set $\mathcal{S} = \emptyset$
\State Factorize the matrix $\mathbf{A}^{-1}$ into $\mathbf{A}^{-1}=\mathbf{U}^{T}\mathbf{U}$.
\State Calculate and save  $\bm{z}_{i}=\mathbf{U}\bm{x}_{i}$ for all $i\in \mathcal{N}$. 
\State For all $e\in \Omega$, insert tuples $(||\bm{z}_{i}-\bm{z}_{j}||_{2}^{2},e)$ into heap $\mathbf{H}$.
\EndProcedure
\end{algorithmic}
\begin{algorithmic}[1]
\Procedure{UpdateMarginal}{$e,S$}
\State \Return $||\bm{z}_{i}-\bm{z}_{j}||_{2}^{2}$ \label{line:facteval}
\EndProcedure
\end{algorithmic}
\begin{algorithmic}[1]
\Procedure{UpdateS}{$\mathcal{S},e^{*}$}
\State $\mathcal{S}=\mathcal{S}\cup e^{*}$; $\mathbf{A}^{-1}=\mathbf{A}^{-1}-\frac{\mathbf{A}^{-1}\bm{x}_{e^{*}}\bm{x}_{e^{*}}^{T}\mathbf{A}^{-1}}{1+\bm{x}_{e^{*}}^{T}\mathbf{A}^{-1}\bm{x}_{e^{*}}}$
\State Factorize the matrix $\mathbf{A}^{-1}$ into $\mathbf{A}^{-1}=\mathbf{U}^{T}\mathbf{U}$.
\State Pre-compute and save  $\bm{z}_{i}=\mathbf{U}\bm{x}_{i}$ for all $i\in \mathcal{N}$. 
\label{line:prec} 
\EndProcedure
\end{algorithmic}
\end{scriptsize}
\end{algorithm}

\noindent\textbf{Factorization Lazy Greedy.} As in Sec.~\ref{subsec:fga}, prior to the loop in \textsc{FindMax} that locates the maximal element, the matrix $\mathbf{A}^{-1}$ can be factorized as $\mathbf{A}^{-1}=\mathbf{U}^T\mathbf{U}$ via Cholesky factorization. Again, vectors $\bm{z}_i=\mathbf{U}\mathbf{x}_i$, $i\in \mathcal{N}$, can be pre-computed and used in subsequent computations of quantities $d_e$ as needed. In theory, as the loop may terminate early, it is best to \emph{not} precompute a vector $\bm{z}_i$, $i\in  \mathcal{N}$, but only compute it the first time some  $e=(i,j)\in\mathcal{C}$, is popped from the heap, and the computation of $d_e$ requires it. Once a $\bm{z}_i$, $i\in \mathcal{N}$ has been computed thusly, it can be re-used again in subsequent comparison pairs $(i,j')\in \mathcal{C}$ that require it. We call this algorithm \emph{Factorization-Lazy-Greedy with Memoization}, as $\bm{z}_i$ computations are \emph{memoized} (i.e., computed as needed and saved to be used later). In practice (see Sec.~\ref{sec:evaluation}),  pre-computing all $\mathbf{z}_i$, $i\in \mathcal{N}$, even if they are not all used in the  subsequent lazy loop evaluation, and paying the corresponding $O(Nd^2)$ cost may be faster when matrix-vector multiplications are optimized; we call this algorithm \emph{Factorization Lazy Greedy with Pre-Computation}. We elaborate on this in Sec.~\ref{sec:evaluation}, where we implement both versions of the algorithm. 

Factorization Lazy Greedy is shown in pseudocode in Alg.~\ref{alg:factorL}; we provide only pseudocode for the pre-computed version. In particular, we pre-compute and save  $\bm{z}_{i}=\mathbf{U}\bm{x}_{i}$ for all $i\in \mathcal{N}$ in line 
\ref{line:prec} of procedure \textsc{UpdateS}, paying an $O(Nd^2)$ cost per iteration. Not all such values are used by the \textsc{Greedy} algorithm however, as a loop may terminate early. In the memoized version, $\bm{z}_{i}$ are computed online/as needed at line \ref{line:facteval} of procedure \textsc{UpdateMarginal}, and stored/reused at later calls.

\begin{algorithm}[!t]
\begin{scriptsize}
\caption{Scalar Lazy Algorithm} \label{alg:scalarL}
Lazy Greedy Algorithm, as described in Sec.~\ref{sec:aclazy}. The main \textsc{Greedy} procedure is  the same as in Alg.~\ref{alg:greedy}.
\begin{algorithmic}[1]
\Procedure {PreProcessing}{$\mathbf{X}$}
\State Compute
	$\mathbf{A}_{0}^{-1}=(\lambda I_{d}+\underset{i\in\mathcal{A}}{\sum}\bm{x}_{i}\bm{x}_{i}^{T})^{-1}$; Set $\mathbf{A}^{-1}=\mathbf{A}_0^{-1}$
\State Factorize the matrix $\mathbf{A}^{-1}$ into $\mathbf{A}^{-1}=\mathbf{U}^{T}\mathbf{U}$.
\State Calculate and save  $\bm{z}_{i}=\mathbf{U}\bm{x}_{i}$ for all $i\in \mathcal{N}$. 
\State For all $e\in \Omega$, insert tuples $(||\bm{z}_{i}-\bm{z}_{j}||_{2}^{2},0,e)$ into heap $\mathbf{H}$.
\State Set $\mathcal{S} = \emptyset, k=0$.
\EndProcedure
\end{algorithmic}
\begin{algorithmic}[1]
\Procedure {FindMax}{$f,\mathcal{S}$}
\While{True}
\State $(\Delta_{\texttt{old}},t,e)=$ $\mathbf{H}$.\textsc{PopMax}()
\State Set $\Delta=\textsc{UpdateMarginal}(\Delta_{old},t,e,S)$
\State $(\Delta',t,e') = \mathbf{H}.$\textsc{Max}()
\If{$\Delta \geq \Delta'$}
\Return e
\Else
\State $\mathbf{H}$.\textsc{Insert}$\big( (\Delta,k,e )\big)$
\EndIf
\EndWhile 
\EndProcedure
\end{algorithmic}
\begin{algorithmic}[1]
\Procedure{UpdateMarginal}{$\Delta_\texttt{old},t,e,S$}
\State \Return $\Delta_\texttt{old}-\underset{t\leqslant l< k }{\sum}(\rho_{l,i}-\rho_{l,j})^{2}$ \label{line:slazupdate}
\EndProcedure
\end{algorithmic}
\begin{algorithmic}[1]
\Procedure{UpdateS}{$\mathcal{S},e^{*}$}
\State Compute and save $\bm{v}_{k}=\frac{\mathbf{A}^{-1}\bm{x}_{e^{*}}}{\sqrt{1+\bm{x}_{e^{*}}^{T}\mathbf{A}^{-1}\bm{x}_{e^{*}}}}$
\State Pre-compute and save  $\rho_{k,i}=\bm{v}_{k}^{T}\bm{x}_{i}$ for all $i\in \mathcal{N}$.
\State $\mathbf{A}^{-1}=\mathbf{A}^{-1}-\bm{v}\bm{v}^{T}$ \label{line:slazprec}
\State $\mathcal{S}=\mathcal{S}\cup e^{*}$	
\State $k=|\mathcal{S}|$
\EndProcedure
\end{algorithmic}
\end{scriptsize}
\end{algorithm}

\noindent\textbf{Scalar Lazy Greedy.} Finally, as in Sec.~\ref{sec:Scalar}, values $d_e$ can be adapted  using formula \eqref{eq:adaptde}. Beyond maintaining and updating the corresponding variables present in Alg.~\ref{alg:sg} ( $z_i=\bm{v}^{T}\mathbf{x}_i$, $i\in\mathcal{N}$, vector $\bm{v}$ given by \eqref{eq:adaptde}, etc.), adapting  $d_e$ via \eqref{eq:adaptde} poses a challenge in the context of lazy greedy: this is because the formula provides the adaptation rule w.r.t. the value $d_e$ in the \emph{immediately preceding iteration}. The values $d_e$ stored and retrieved (via a pop) from the heap may have been computed at an arbitrarily old iteration. Hence, to construct the (approximate) marginal gain $d_e$ under the current set $S$ from a popped value from the heap we may need to repeatedly apply \eqref{eq:adaptde} more than once. This  requires to also keep track the iteration at which tuples are inserted in the heap, so that the appropriate vectors $\bm{v}$ can be used to adapt them. We indeed track this information in Alg.~\ref{alg:scalarL}.

We note that, as a result, the execution of \textsc{UpdateMarginal} may be quite expensive when popped values of the heap are quite ``stale'' (i.e., were computed in very early iterations). Hence, in contrast to the standard greedy versions of these algorithms, it is not a-priori obvious that Scalar Lazy Greedy always outperforms Factorization Lazy Greedy; we indeed observe the opposite in our experiments in Sec.~\ref{sec:evaluation}. Finally, as in Factorization Lazy Greedy, multiplications $\bm{v}^T\bm{x}_i$ can again either be fully pre-computed at each iteration (paying the full $O(Nd)$ cost), or memoized and used as necessary. We again implement and evaluate both options in Sec.~\ref{sec:evaluation}. 

Pseudocode for Scalar Lazy Greedy can be found in  Alg.~\ref{alg:scalarL}. Again, we provide pseudocode only  only for the version that uses pre-computation. In particular,
 quantities $\rho_{k,i}$, for $k\in\{1,\ldots,K\}$, $i\in\mathcal{N}$, are pre-computed at line \ref{line:slazprec} of \textsc{UpdateS}; in a memoized version, they can again be computed online/as needed at line \ref{line:slazupdate} of \textsc{UpdateMarginal}. Note that, in both cases, this requires computing and saving vectors $\bm{v}_k$, that can be used as necessary.

%
%

\section{Real Datasets}\label{sec:AppendixB}
We provide here a detailed description of the datasets we use in our experiments.

\noindent\emph{ROP Dataset.}
Our first dataset \cite{kalpathy2016plus} consists of 100 images of retinas, labeled by experts w.r.t. the presence of a disease called
Retinopathy of Prematurity (ROP).
We represent each image through a vector $\bm{x}_i\in \mathbb{R}^d$ where $d=156$, using the feature extraction procedure of \cite{kalpathy2016plus}, comprising statistics of several indices such as blood vessel curvature, dilation, and tortuosity. Five experts provide diagnostic labels for all 100 images, categorizing them as Plus, Preplus and Normal. We convert these to absolute labels $y_i\in\{-1,+1\}$ by mapping Plus and Preplus as $+1$  and Normal to $-1$.
Finally, these five experts also provide $|\mathcal{C}|=29705$ comparison labels for 4950 pairs of images in this dataset. 
Beyond these labels, we also have Reference Standard Diagnosis (RSD) labels for 
each of these images, which are created via a consensus reached by a committee of 3 experts. We use these additional labels for testing purposes, as described below.

\noindent\emph{ROP5K Dataset.}
The ROP5K dataset \cite{brown2018fully} consists of $N=5000$ unlabeled images of retinas.  Each image has a feature dimension $d=143$, generated again via the feature extraction process of \cite{kalpathy2016plus}. We execute 150 experiments on random samples of size $N=3000$ from this dataset, and report performance averages.

\noindent\emph{SUSHI Dataset.}
The SUSHI Preference dataset \cite{kamishima2009trbagg} consists of rankings of $N=100$ sushi food items by 5000 customers.  Each customer ranks 10 items according to her preferences.  Each sushi item is associated with a feature vector $\bm{x}_{i}\in \mathbb{R}^{d}$ where $d=20$, consisting of features such as style, group,  heaviness/oiliness in taste, frequency, and normalized price.
We generate comparison labels as follows. For any pair of items $i,j\in \mathcal{N}$ in a customer's ranked list, if $i$ precedes $j$ in the list, we set $y_{i,j}=+1$, otherwise, $y_{i,j}=-1$. We also produce absolute labels via the Elo ranking algorithm \cite{elo1978rating}. This gives us an individual score for each item; 
we convert the individual score to an absolute label $y_i\in\{-1,+1\}$ by setting items above (below) the median score to $+1$ ($-1$).

\noindent\emph{Netflix Dataset.}
The Netflix dataset has multiple users and  $17770$ movies. We select 150 users who have rated more than 833 movies. Each movie has a 30-dimensional vector obtained via matrix factorization \cite{koren2009matrix} over the entire dataset. We generate binary absolute labels as follows: if the rate score  is above (below) the user mean, the absolute label is $+1$ ($-1$). If the scores between two movies $i$ and $j$ are different, we generate comparison label $y_{i,j}=+1$ if score $i$ is higher than score $j$, otherwise we break ties by setting $y_{i,j}=-1$.


\noindent\emph{MSLR Dataset.}
The  MSLR-WEB10K dataset \cite{DBLP:journals/corr/QinL13} has $10000$ queries. The datasets consist of $134$-dimensional features such as covered query term number, covered query term ratio, stream length, inverse document frequency (IDF), 
etc. We restrict the dataset to 150 queries submitted more than 325 times.

\noindent\emph{SIFT dataset.}
The SIFT10M dataset \cite{Dua:2017} has often been used for evaluating the approximate nearest neighbour search methods. Each data point is a SIFT feature which is extracted from Caltech-$256$ by the open source VLFeat library. The dataset has a total of 11164866 instances and  each SIFT feature has a dimensionality of 128. We execute 150 experiments on random samples of size $N=3000$ in this dataset, and report performance averages.

\noindent\emph{CAMRa Dataset.}
The CAMRa dataset \cite{bento2011identifying} has multiple users and  $23893$ movies. We select 150 users who have rated more than 896 movies. Each movie has a 10-dimensional vector obtained via matrix factorization over the entire dataset.

\section{Competitor Methods}\label{sec:b2}

We implement the greedy algorithm with the following objectives (see also~\cite{guo}):

\noindent\textbf{Mutual Information.}
Recall that the prior distribution is $\bm{\beta} \sim \mathcal{N}(0,\sigma^2 I_{d})$. The objective function is to maximize the mutual information between the parameter vector $\bm{\beta}$ and  selected comparison labels $Y_{\mathcal{S}}$, conditioned on the observed absolute labels, i.e:
\begin{equation}\label{eq5}
\begin{split}
    f_{1}(\mathcal{S})=&\mathbf{I}(\bm{\beta};Y_{\mathcal{S}}|Y_{\mathcal{A}}=y_{\mathcal{A}})\\
    =&\mathbf{H}(Y_{\mathcal{S}}|Y_{\mathcal{A}}=y_{\mathcal{A}})-\mathbf{H}(Y_{\mathcal{S}}|\bm{\beta},Y_{\mathcal{A}}=y_{\mathcal{A}}),
\end{split}
\end{equation}
where $\mathbf{I}(\cdot|Y_{\mathcal{A}}=y_{\mathcal{A}})$ denotes the mutual information conditioned on the observed absolute labels and $\mathbf{H}(\cdot|Y_{\mathcal{A}}=y_{\mathcal{A}})$ denotes the entropy conditioned on the observed absolute labels. We compute the quantities in Eq.~\eqref{eq5} using the Bradley-Terry generative model described in \eqref{eq2}.

\noindent\textbf{Information Entropy.}
Recall that given some observed absolute labels $y_{\mathcal{A}}$, we can estimate the parameter vector  $\hat{\bm{\beta}}$ by:
\begin{equation}\label{beta}
\begin{split}
\hat{\bm{\beta}}=\textstyle\mathop{\mathbf{argmax}}_{\bm{\beta}} \mathcal{L}(\bm{\beta};\mathcal{A},\emptyset),
\end{split}
\end{equation}
where the negative log-likelihood function  $\mathcal{L}(\bm{\beta};\mathcal{A},\mathcal{S})$ is given by Eq.~\eqref{eq4}.
Under our generative model, unlabeled samples are independent given $\hat{\bm{\beta}}$; hence, the information entropy objective  can be written as:
\begin{align}
f_{2}(\mathcal{S})= \textstyle\mathbf{H}(Y_{\mathcal{S}}|\bm{\beta}=\hat{\bm{\beta}})=\sum_{a\in\mathcal{S}}\mathbf{H}(Y_{a}|\bm{\beta}=\hat{\bm{\beta}}).
\end{align}
Assuming that the experimenter estimates the parameter vector $\hat{\bm{\beta}}$, thus we can use information entropy to measure the unpredictability of $Y_{\mathcal{S}}$ . This can be seen as a ``point'' estimate of the mutual information. 

\noindent\textbf{Fisher Information.}
The Fisher information measures the amount of information that an observable random feature $\bm{x}$ carries about an unknown parameter $\bm{\beta}$ upon which the probability of $\bm{x}$ depends.
The Fisher information matrix can be written as:
\begin{equation}\label{fish}
    \textstyle I(\bm{\beta})=-\int p(y|\bm{x},\bm{\beta})\frac{\partial^2 }{\partial \bm{\beta}^2} \mathop{\log}p(y|\bm{x},\bm{\beta})\mathrm{d}\bm{x}\mathrm{d}y.
\end{equation}
Let $p(\bm{x})$ be the feature distribution of all unlabeled examples in set $\mathcal{C}$ and $q(\bm{x})$ be the distribution of unlabeled examples in set $\mathcal{S}$ that are chosen for manual labeling. With the generative model and the estimation of parameter vector $\hat{\bm{\beta}}$ by Eq.~\eqref{beta}, the Fisher information matrices for these two distributions can be written as:
\begin{small}
\begin{align*}
\begin{split}
I_{p}(\hat{\bm{\beta}})&=\textstyle\frac{1}{|\mathcal{C}|}{\sum}_{(i,j)\in\mathcal{C}}\pi(\bm{x}_{i,j})(1-\pi(\bm{x}_{i,j}))\bm{x}_{i,j}\bm{x}_{i,j}^{T}+\delta I_{d},\\
I_{q}(\mathcal{S},\hat{\bm{\beta}})&=\textstyle\frac{1}{|\mathcal{S}|}{\sum}_{(i,j)\in\mathcal{C}}\pi(\bm{x}_{i,j})(1-\pi(\bm{x}_{i,j}))\bm{x}_{i,j}\bm{x}_{i,j}^{T}+\delta I_{d},
\end{split}
\end{align*}
\end{small}
\vskip 0.0 in
\noindent
where $\delta\ll1$ is to avoid  having a singular matrix,  \\
 $   \pi(\bm{x}_{i,j})=\frac{1}{1+\mathop{\exp}(-\bm{\beta}^{T}\bm{x}_{i,j})},$ for  $(i,j)\in \mathcal{C}$, and $\bm{x}_{i,j}=\bm{x}_i-\bm{x}_j$.
 The matrices above relate to variance of the parameter estimate via the so-called Cramer-Rao bound maximizing
\begin{equation}\label{eqfish}
f_{4}(\mathcal{S})=-\mathop{\mathrm{tr}}(I_{q}(\mathcal{S},\hat{\bm{\beta}})^{-1}I_{p}(\hat{\bm{\beta}}))
\end{equation}
minimizes the Cramer-Rao bound of the respective  $\hat{\bm\beta}$.

\section{Accuracy $\&$ Time Efficiency vs Competitors}\label{sec:AppendixD}
Here, we provide the accuracy and time efficiency result for the Sushi, Netflix and Camra datasets. For Figure. \ref{fig:aucsushi} to \ref{fig:auccam}, we show the accuracy and time efficiency for both D-optimal design and competitors on Sushi, Netflix and Camra datasets. 
We reach the same conclusion as the ones reported in Sec. \ref{sec:evaluation}. 
We note that, for the Netflix and Camra datasets, we can only compute a batch size less than ten for Fisher Information and Mutual Information. This is because the Mutual has a complexity $O(N^2 2^{K})$ and Fisher has a complexity $O(N^{4}K)$. In the Sushi and Netflix dataset, the best AUC comes from the D-optimal design algorithm. For the Camra dataset, as the dimension is only ten, the D-optimal design algorithm can only beat Random when the batch size is less than 100. The Entropy method is time efficient but has worse accuracy than other methods. The D-optimal Naive Greedy and its variant have good accuracy and are time efficient and after the acceleration FLP and SLP are even faster than Entropy method.

\begin{figure}[!t]
\centering
\includegraphics[width=1.0\columnwidth, height=0.55\columnwidth]{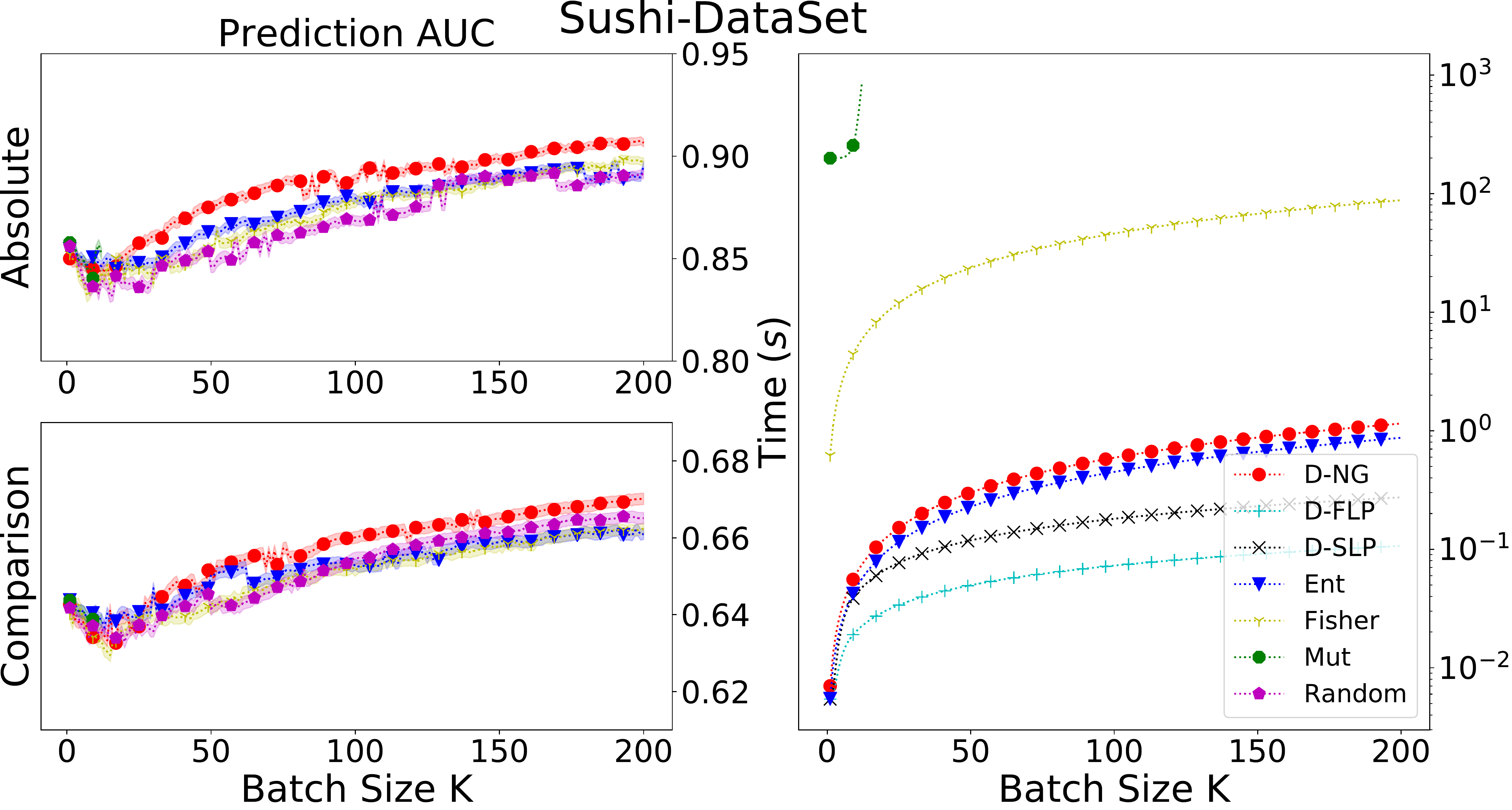}
\caption{Test set AUC and execution time for Sushi dataset, when comparisons samples are selected via D-optimal, Fisher, Entropy and Random and Mutual Information. The classifier is trained via MAP \eqref{eq4} on the training set.  The left figure is the test AUC for absolute label, the middle figure is the test AUC for comparison labels. The right figure is the execution time for different algorithms. Especially, for the D-optimal method we record the execution time for Naive Greedy, Factorization Lazy and Scalar Lazy Greedy Algorithm. For Mutual information we only execute it for batch size no more than 12.}
\label{fig:aucsushi}
\end{figure}

\begin{figure}[!t]
\centering
\includegraphics[width=1.0\columnwidth, height=0.55\columnwidth]{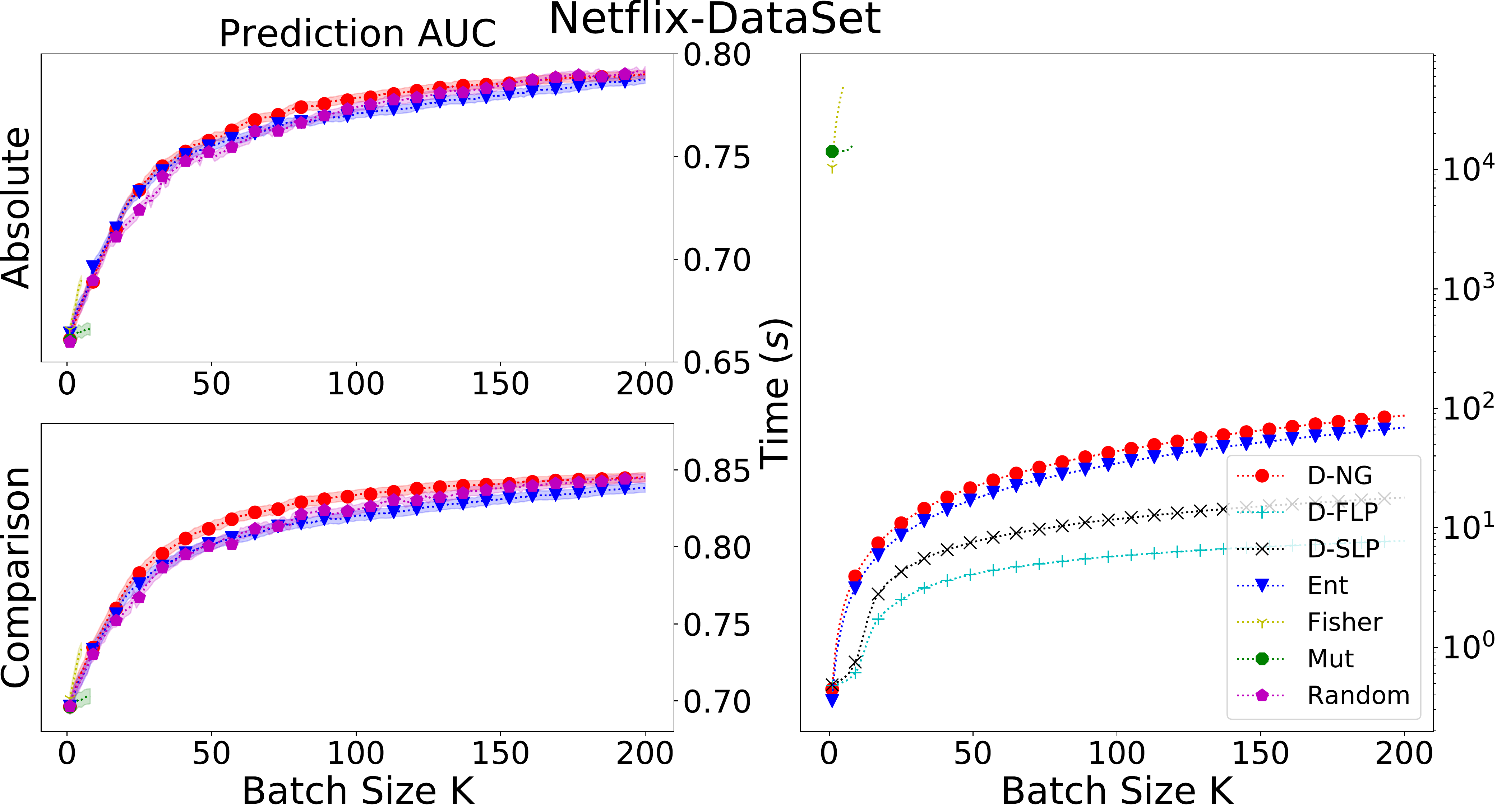}
\caption{Test set AUC and execution time for Netflix dataset, when comparisons samples are selected via D-optimal, Fisher, Entropy and Random and Mutual Information. The classifier is trained via MAP \eqref{eq4} on the training set.  The left figure is the test AUC for absolute label, the middle figure is the test AUC for comparison labels. The right figure is the execution time for different algorithms. Especially, for the D-optimal method we record the execution time for Naive Greedy, Factorization Lazy and Scalar Lazy Greedy Algorithm. For Mutual information we only execute it for batch size no more than 8, for Fisher information method we only execute it for batch size no more than 5.}
\label{fig:aucnet}
\end{figure}

\begin{figure}[!t]
\centering
\includegraphics[width=1.0\columnwidth, height=0.5\columnwidth]{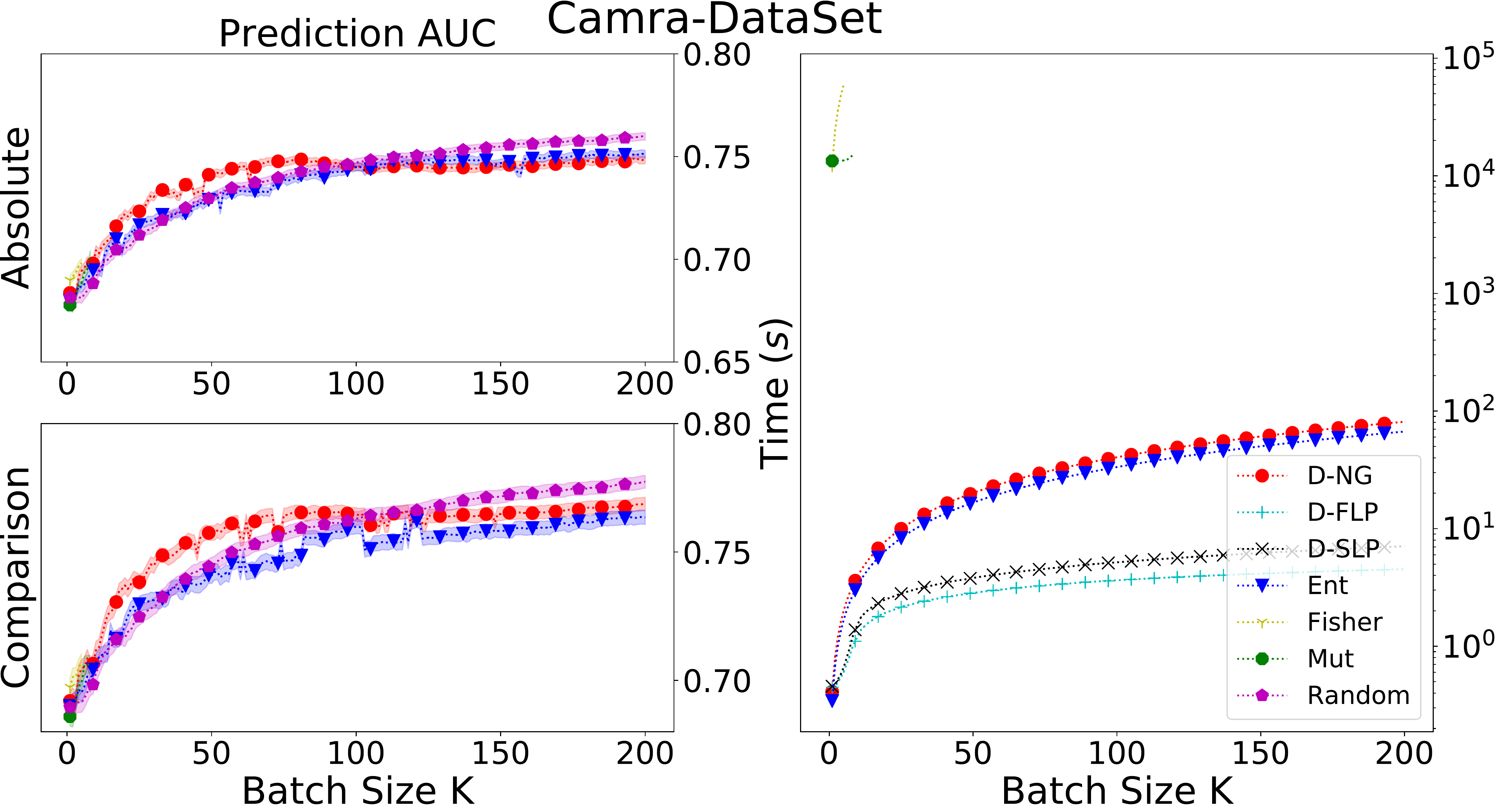}
\caption{Test set AUC and execution time for Camra dataset, when comparisons samples are selected via D-optimal, Fisher, Entropy and Random and Mutual Information. The classifier is trained via MAP \eqref{eq4} on the training set.  The left figure is the test AUC for absolute label, the middle figure is the test AUC for comparison labels. The right figure is the execution time for different algorithms. Especially, for the D-optimal method we record the execution time for Naive Greedy, Factorization Lazy and Scalar Lazy Greedy Algorithm. For Mutual information we only execute it for batch size no more than 8, for Fisher information method we only execute it for batch size no more than 5.}
\label{fig:auccam}
\end{figure}
}

\end{document}